\documentclass[aps,pr,superscriptaddress,preprintnumbers,nofootinbib,10pt]{revtex4-2}

\usepackage{multirow}
\usepackage{amsmath}
\usepackage{amssymb}
\usepackage[dvipdf,dvips]{graphicx}
\usepackage{color}
\usepackage{hyperref}
\usepackage{url}
\usepackage{slashed}
\usepackage{subfigure}
\usepackage[usenames,dvipsnames]{xcolor}
\usepackage{amsmath}
\usepackage{amsfonts}
\usepackage{float} 
\usepackage{amssymb}
\usepackage{epsfig}
\usepackage{graphics}
\usepackage{euscript}
\usepackage{slashed}
\usepackage{epstopdf}
\usepackage[utf8]{inputenc}
\allowdisplaybreaks
\usepackage[normalem]{ulem}
\usepackage{pifont}
\usepackage{dsfont}
\usepackage{MnSymbol}
\usepackage{verbatim}
\usepackage{graphicx}
\usepackage{latexsym}
\usepackage{tikz-feynman}
\usepackage{bbold}

\begin{document}

	\title{{\bf \Large Bell's inequality in relativistic Quantum Field Theory }}
	
	\vspace{1cm}
	
	\author{M. S.  Guimaraes}\email{msguimaraes@uerj.br} \affiliation{UERJ $–$ Universidade do Estado do Rio de Janeiro,	Instituto de Física $–$ Departamento de Física Teórica $–$ Rua São Francisco Xavier 524, 20550-013, Maracanã, Rio de Janeiro, Brazil}
	
	\author{I. Roditi} \email{roditi@cbpf.br} \affiliation{CBPF $-$ Centro Brasileiro de Pesquisas Físicas, Rua Dr. Xavier Sigaud 150, 22290-180, Rio de Janeiro, Brazil } \affiliation{Institute for Theoretical Physics, ETH Zürich, 8093 Zürich, Switzerland} 
	
	\author{S. P. Sorella} \email{silvio.sorella@fis.uerj.br} \affiliation{UERJ $–$ Universidade do Estado do Rio de Janeiro,	Instituto de Física $–$ Departamento de Física Teórica $–$ Rua São Francisco Xavier 524, 20550-013, Maracanã, Rio de Janeiro, Brazil}

	\begin{abstract}

A concise and self-contained introduction to the Bell inequality in relativistic Quantum Field Theory is presented. Taking the example of a real scalar massive field, the violation of the Bell inequality in the vacuum state and for causal complementary wedges is illustrated. 
		
		\end{abstract}

	\maketitle

	\vspace{1cm}
	
	\tableofcontents

	\section{Introduction}\label{Sect1}
	
Since its discovery \cite{Bell:1964kc}, the Bell inequality is a continuous source of investigation with astounding applications in many areas such as: Quantum computation, Quantum teleportation, Quantum cryptography, to mention a few of them. \\\\Although being devoted to non-relativistic Quantum Mechanics, there are no doubts that the concepts of the Minkowski spacetime and of the relativistic causality permeate all  Bell's work as well as the seminal paper by Einstein-Podolsky-Rosen \cite{Einstein:1935rr}. It seems thus that the most natural step is to move ahead and consider the Bell inequality within the realm of relativistic Quantum Field Theory, an exceptional setup unifying Quantum Mechanics and Special Relativity. Needless to say, in a relativistic Quantum Field Theory, all causality issues related to the Minkowski structure of the spacetime are manifestly present at all stages. \\\\One could expect that, in  Quantum Field Theory, the Bell inequality might exhibit both a simple formulation as well as a concrete computational framework. Unfortunately, life is not that easy. Despite the remarkable results achieved in the eighties and in the nineties \cite{Summers:1987fn,Summ,Summers:1987ze}, the study of the Bell inequlity in Quantum Field Theory remains  a big challenge. \\\\On one hand,  the construction of the field correlation functions and the identification of the regions in Minkowski space leading to a violation of the Bell inequality require the use of  powerful results which have demanded a long and deep investigation on the foundations of the Quantum Theory and of the nature of the spacetime. We might quote:  the von Neumann algebras \cite{BR}, the algebraic formulation of the Quantum Field Theory as encoded in the Haag-Kastler approach 
\cite{Haag:1992hx}, the  modular theory of Tomita-Takesaki \cite{BR,Summers:2003tf,Guido:2008jk}, the Reeh-Schlieder \cite{Reeh:1961ujh} and the Bisognano-Wichmann \cite{BW} Theorems.  Nowadays, all these tools are  part of the general setup at disposal of field theorists aiming at unravelling the intricacies of the entanglement, unavoidably present in a relativistic Quantum Field Theory \cite{Witten:2018zxz}. \\\\On the other hand, it seems safe to state that the explicit construction of the bounded  operators entering the Bell inequality and the effective computation of their correlation functions still need much work. \\\\In these notes we aim at presenting, in a self-contained and sequential fashion, all the aforementioned topics, the goal being that of providing an updated summary of the Bell inequality in Quantum Field Theory. Moreover, taking as explicit example the case of a real massive free scalar field, we shall illustrate how a certain class of correlation functions can be concretely evaluated, leading to sensible violations of the Bell inequality. \\\\It is worth underlining  that, besides the theoretical studies, the Bell inequality is receiving great attention in high energy particle physics, both at phenomenological and experimental level, see the recent review  \cite{Barr:2024djo}.\\\\Here, we are particularly interested in the Bell-CHSH scenario, where two distant observers,  very often named Alice and Bob, perform local measurements on entangled subsystems. Each observer is allowed to choose between two measurement settings. The corresponding correlation functions are then combined into the relation,
\[
{\cal C} = E(A_1,B_1) + E(A_1,B_2) + E(A_2,B_1) - E(A_2,B_2),
\]
where \(E(A_i,B_j)\) denotes the expectation value of the product of the outcomes when measurement settings \(A_i\) and \(B_j\) are chosen. Local realistic theories impose the bound \(|{\cal C}| \leq 2\); however, quantum mechanics predicts violations up to Tsirelson's bound, \(|{\cal C}| \leq 2\sqrt{2}\). This discrepancy forms one of the most striking demonstrations of quantum nonlocality. In the framework of Quantum Field Theory, the challenge is to construct the corresponding bounded operators and rigorously compute their correlation functions, taking into account the intrinsic requirements of relativistic causality. \\\\In the context of Algebraic Quantum Field Theory, the task is to construct the observables \(A_i\) and \(B_j\) as bounded operators on the Hilbert space, using the algebra of local observables associated with regions in Minkowski space. The algebraic approach provides a rigorous framework to deal with these issues, as it naturally incorporates locality and causality through the Haag-Kastler axioms and facilitates the description of entangled states via the structure of von Neumann algebras. \\\\The pioneering work by Summers and Werner \cite{Summers:1987fn,Summ,Summers:1987ze} demonstrated that for certain choices of observables localized in complementary wedge regions, the vacuum state or, more generally, any state satisfying the spectrum condition and modular properties, exhibits a violation of the Bell-CHSH inequality. Their analysis makes extensive use of modular theory, particularly the Tomita-Takesaki theory, to rigorously define the local algebras and to analyze the properties of the vacuum state. Moreover, these works highlight that the nonlocal correlations intrinsic to the vacuum of a Quantum Field Theory can be detected by appropriately constructed bounded operators even though the individual field operators are usually unbounded. \\\\Furthermore, the algebraic formulation has the advantage of being model-independent. That is, one can derive general results regarding the violation of Bell type inequalities without relying on the detailed dynamics of the fields; rather, the properties of the underlying local algebras and their modular structure are sufficient to produce these violations. This line of research not only confirms the robustness of quantum correlations but also opens up new avenues for exploring the interplay between entanglement, locality, and relativistic causality in Quantum Field Theory.  \\\\The material has been organized as follows. In Sect.\eqref{Sect2} we review the canonical quantization of the free real scalar field and the construction of the corresponding Hilbert space. Sect.\eqref{Vn} contains a concise account on the von Neumann algebras and their classification. Sect.\eqref{HK} is devoted to the presentation of few elements of the Algebraic Quantum Field Theory: the Haag-Kastler construction, the Reeh-Schlieder and the BIsognano-Wichmann theorems, the Haag duality.  In Sect.\eqref{Sect6} we discuss the Quantum Field Theory formulation of the Bell inequality. More precisely, we shall focus on a version of the Bell inequality, commonly referred as the Bell-CHSH inequality \cite{Clauser:1969ny}. We shall present the results obtained by Summers-Werner \cite{Summers:1987fn,Summ,Summers:1987ze}, specializing to the case of the vacuum state $|0\rangle$ and for causal complementary wedges regions in spacetime. As already mentioned, we shall also construct a set of correlation functions built out with Weyl operators which can be concretely evaluated. The  characterization  of these correlation functions will be addressed also in  Sect.\eqref{na}, where a numerical approach for wedge regions will be outlined, leading to explicit violations of the Bell-CHSH inequality. In Sect.\eqref{Cc}  we collect  our conclusion, illustrating in a certainly partial and incomplete way   the huge amount of work which remains to be done  to master the Bell-CHSH in Quantum Field Theory. \\\\Let us end with a few words on the level of these notes. As the reader can easily figure out, the task of presenting a  self contained account  on the Bell inequality in Quantum Field Theory is highly complex. A mathematical rigorous presentation of all needed issues would have led to a very long unbounded review.  We have adopted therefore the strategy of providing brief accounts focusing on the illustration of the main concepts and results, reminding the interested reader to the more specific literature. As such, we are indebted with our mathematically oriented colleagues. Though, we hope that the material will be helpful to students and researchers willing to face  this beautiful and fascinating challenging topic.

\section{Canonical quantization  of the free real scalar field}\label{Sect2}
	
Let us start by reviewing   the  canonical quantization of the free real massive scalar field  in $1+3$ Minkowski spacetime  \cite{Haag:1992hx}, with signature $diag(+,-, -,-)$:  
\begin{equation} 
{\cal L} =   \frac{1}{2} \left( \partial^\mu \varphi \partial_\mu \varphi - m^2 \varphi^2 \right).  \label{cnq1}
\end{equation} 
From the field equation 
\begin{equation} 
(\partial^\mu \partial_\mu + m^2) \varphi(t,{\vec x})= 0 \;, \label{eqmvphi} 
\end{equation} 
one gets the plane-wave expansion 
\begin{equation} \label{qf}
\varphi(t,{\vec x}) = \int \frac{d^3 {\vec k}}{(2 \pi)^3} \frac{1}{2 \omega_k} \left( e^{-ikx} a_k + e^{ikx} a^{\dagger}_k \right), 
\end{equation} 
where $\omega_k  = k^0 = \sqrt{{\vec{k}}^2 + m^2}$. The canonical quantization of $\varphi(t,{\vec x})$ is implemented by promoting $(a_k, a^{\dagger}_k)$ 
to annihilation and creation  operators obeying the canonical commutation relations
\begin{align}\label{ccr}
[a_k, a^{\dagger}_q] &= (2\pi)^3 2\omega_k \delta^3({\vec{k} - \vec{q}}), \\ \nonumber 
[a_k, a_q] &= [a^{\dagger}_k, a^{\dagger}_q] = 0, 
\end{align}
A simple calculation yields
\begin{equation}
\left[ \varphi(x) , \varphi(y) \right] = i \Delta_{\textrm{PJ}} (x-y), \label{caus} 
\end{equation}
where $\Delta_{\textrm{PJ}}(x-y) $ is the Lorentz-invariant causal Pauli-Jordan function, defined by \footnote{$\varepsilon(x) \equiv \theta(x) - \theta(-x)$}
\begin{align}\label{it1}
	 i \Delta_{\textrm{PJ}}(x-y) \!=\!\! \int \!\! \frac{d^4k}{(2\pi)^3} \varepsilon(k^0) \delta(k^2-m^2) e^{-ik(x-y)}. 
\end{align}
The Pauli-Jordan function vanishes when $(x-y)^2 < 0$, thus encoding the principle of relativistic causality. Furthermore, it  satisfies $(\partial^2_x + m^2) \Delta_{\textrm{PJ}}(x-y) = 0$ and $\Delta_{\textrm{PJ}}(x-y) = - \Delta_\textrm{{PJ}}(y-x)$. Explicitly, $ \Delta_{\textrm{PJ}}(x-y)$  can be written as
 \begin{equation} 
 	\Delta_{\textrm{PJ}}(x-y) =\frac{\varepsilon(x^0-y^0) }{2\pi}  \left(  -\delta\left((x-y)^2\right)  
 	  +m \frac{\theta((x-y)^2)  J_1(m\sqrt{(x-y)^2}) }{ 2 \sqrt{(x-y)^2}}\right) \;,
 \end{equation} 
 where $J_1$ is the Bessel function. \\\\The quantum fields are too singular objects, being in fact operator-valued distributions~\cite{Haag:1992hx}. As such, they have to be smeared out in order to provide well defined operators acting in the Hilbert space, namely  
\begin{equation} \label{sm}
\varphi(h) = \int d^4x \;\varphi(x) h(x) \;. 
\end{equation} 
The singular nature of $\varphi(x)$ requires that $h(x)$ be a smooth function belonging to the Schwartz space ${\cal S}(\mathbb{R}^4)$, {\it i.e.} to the space of infinitely differentiable continuous functions which go to zero at infinity faster than any power of $x$ in every spacetime direction. A subspace of ${\cal S}(\mathbb{R}^4)$ which will be frequently employed is that of the smooth functions with compact support, called the space of the test functions and denoted by ${\cal C}_{0}^{\infty}(\mathbb{R}^4)$. \\\\Moving to momentum space 
\begin{equation} 
{\hat h}(p) = \int d^4x \; e^{ipx} h(x) \;,  \label{msp}
\end{equation} 
and plugging expression~\eqref{qf} in~\eqref{sm}, one finds 
\begin{equation} \label{smft}
\varphi(h) = \int \frac{d^3 {\vec k}}{(2 \pi)^3} \frac{1}{2 \omega_k} \left( {\hat h}^{*}(\omega_k,{\vec k}) a_k + {\hat h}(\omega_k,{\vec k}) a^{\dagger}_k \right) \;, 
\end{equation}
namely
\begin{align}\label{key}
	\varphi(h) = a_h + a^{\dagger}_h,
\end{align}
where the smeared  creation and annihilation operators are  defined as
\begin{align} 
a_h &= \int \frac{d^3 {\vec k}}{(2 \pi)^3} \frac{1}{2 \omega_k}  {\hat h}^{*}(\omega_k,{\vec k}) a_k, \nonumber \\
a^{\dagger}_h &= \int \frac{d^3 {\vec k}}{(2 \pi)^3} \frac{1}{2 \omega_k} {\hat h}(\omega_k,{\vec k}) a^{\dagger}_k \;. \label{aad} 
\end{align} 
The canonical commutation relations now read
\begin{equation} 
\left[ a_h, a^{\dagger}_{h'}\right]  = \langle{  h} \vert { h}' \rangle \;, \qquad \left[ a_h, a_{h'}\right]=0 \;, \qquad \left[ a^{\dagger}_h, a^{\dagger}_{h'}\right] =0 \;,  \label{ccrfg}
\end{equation}
where $ \langle { h} \vert { h}' \rangle$ stands for the Lorentz invariant scalar product between  ${ h}$ and ${ h}'$:
\begin{equation} 
\langle { h} \vert { h}' \rangle = \int \frac{d^3 {\vec k}}{(2 \pi)^3} \frac{1}{2 \omega_k}  {\hat h}^{*}(\omega_k,{\vec k}) {\hat h'}(\omega_k, {\vec k}) 
= 
\int \frac{d^4 {\vec k}}{(2 \pi)^4} 2\pi \;\theta(k^0) \delta(k^2-m^2)  {\hat h}^{*}(k) {\hat h}'(k). \label{scpd}
\end{equation} 
The scalar product  \eqref{scpd} can be rewritten in configuration space as
\begin{equation} 
\langle { h} \vert { h}' \rangle = \int d^4x d^4x'\; h(x) {\cal D}(x-x') h'(x'), \label{confsp} 
\end{equation} 
where ${\cal D}(x-x')$ is the so-called Wightman function \cite{Streater:1989vi}
\begin{equation} 
{\cal D}(x-x') = \langle 0 \vert \varphi(x) \varphi(x')  \vert 0 \rangle \!=\!\!\! \int \!\!\! \frac{d^3 {\vec k}}{(2 \pi)^3} \frac{1}{2 \omega_k} e^{-ik(x-x')}\;, \label{Wg}
\end{equation} 
where $|0\rangle$ is the vacuum state, $a_p|0\rangle =0$ for all $p$. Expression \eqref{Wg} 
can be decomposed as 
\begin{equation} 
{\cal D}(x-x') = \frac{i}{2} \Delta_{\textrm{PJ}}(x-x')   + H(x-x'), \label{decomp}
\end{equation} 
where $H(x-x')=H(x'-x)$, called the Hadamard function,  is the real symmetric quantity
\begin{equation} 
H(x-x') = \frac{1}{2} \int \frac{d^3 {\vec k}}{(2 \pi)^3} \frac{1}{2 \omega_k} \left( e^{-ik(x-x')} + e^{ik(x-x')}
 \right). \label{H}
\end{equation} 
The commutation relation~\eqref{caus} can be expressed in terms of the smeared quantum fields as
\begin{equation} 
\left[ \varphi(h) , \varphi(h') \right] =i  \Delta_{\textrm{PJ}}(h,h'), \label{comm_smeared}
\end{equation}
where $h$, $h'$ are test functions and
\begin{equation} 
\Delta_{\textrm{PJ}}(h,h')= \int d^4x \; d^4x' h(x) \Delta_{\textrm{PJ}}(x-x')  h'(x'). \label{pauli_jordan_smeared}
\end{equation}
Therefore, the causality condition in terms of smeared fields can be recast as: if the supports of $h$ and $h'$ are space-like separated, then  $\left[ \varphi(h) , \varphi(h') \right] = 0$. \\\\Let us end this section by reminding how the field $\varphi(x)$  transforms under Poincar{\'e} transformations
\begin{equation} 
x'= \Lambda x + a \;, \qquad a \in {\mathbb{R}^4} \;, 
\end{equation}
with $\Lambda$ being an   orthochronous Lorentz transformation \cite{Scharf:1996zi}. One has 
\begin{equation} 
\varphi(\Lambda x + a) = {\cal U}(\Lambda,a) \;\varphi(x) \; {\cal U}^{-1}(\Lambda,a) \;, \label{LU}
\end{equation}
where ${\cal U}(\Lambda,a)$ denotes a unitary representation of the Poincar{\'e} group. 

\subsection{The Hilbert space}\label{Hilb}

Let us turn to the construction of the Hilbert space of the real free scalar field. The usual way of presenting this topic is through the definition  of the Fock space \cite{Scharf:1996zi,Guido:2008jk}.

\subsubsection{The Fock space}  \label{fock1}
One starts by introducing the vacuum state $|0\rangle$, left invariant by Poincar{\'e} transformations, namely 
\begin{equation}
{\cal U}(\Lambda,a) |0\rangle = |0\rangle \;, \label{Pi}
\end{equation}
with 
\begin{equation} 
a_p |0\rangle =0 \;\;\;\forall \;p \;, \qquad \langle 0 |0\rangle =1 \;. \label{vc}
\end{equation} 
One proceeds by introducing the so called 1-particle space ${\cal H}_1$. Following \cite{Guido:2008jk}, it is helpful to introduce the embedding 
\begin{equation} 
h(x) \in {\cal S}(\mathbb{R}^4) \rightarrow {\hat h}(p) \in L^2(H_m, d \mu_m)  \;, \label{em}
\end{equation}where $ d \mu_m(p)$ stands for 
\begin{equation} 
d \mu_m(p)= \frac{d^3 {\vec p}}{(2 \pi)^3} \frac{1}{2 \omega_p} \;, \label{dmu}
\end{equation}
and $p \in H_m= \{ p; \; p^2=m^2,\; p_0>0\}$. As in eq.\eqref{msp}, the quantity ${\hat h}(p)$  is the Fourier transformation of $h(x)$. It should be kept in mind that, although $h(x)$ is real, its Fourier transform is complex
\begin{equation} 
{\hat h}(p) = {\hat h}_1(p) + i {\hat h}_2(p) \;. \label{ch}
\end{equation}
The space $L^2(H_m, d \mu_m)$ is the complex Hilbert space of the square integrable functions ${\hat h}(p)$ equipped with the Lorentz invariant scalar product, eq.\eqref{scpd}, 
\begin{equation} 
\langle {h} \vert { h}' \rangle = \int d\mu_m(p)\; {\hat h}^{*}(p) {\hat h'}(p)  \;. \label{scpd1}
\end{equation} 
For the norm of ${ h}$, we have 
\begin{equation}  
|| { h} ||^2 =  \int d\mu_m(p)\; {\hat h}^{*}(p) {\hat h}(p)  \;< \;\infty \;. \label{nscpd}
\end{equation}
The 1-particle  space ${\cal H}_1$ is the set of all vectors 
\begin{equation} 
{\cal H}_1 = \{ a^{\dagger}_h |0\rangle, \; h\in {\cal S}(\mathbb{R}^4) \}  \;, \label{H1}
\end{equation}
where the smeared creation operator $a^{\dagger}_h$ has been defined in eq.\eqref{aad}. For the scalar product between two vectors of ${\cal H}_1$ one has 
\begin{equation} 
\langle 0|\; a_h a^{\dagger}_g \; |0\rangle = \int d\mu_m(p)\; {\hat h}^{*}(p) {\hat g}(p)  = \langle h|g\rangle \;. \label{is}
\end{equation}
This important relation shows that the 1-particle space is isomorphic to the Hilbert space $L^2(H_m, d \mu_m)$ \cite{Scharf:1996zi,Guido:2008jk}: 
\begin{equation} 
{\cal H}_1 = L^2(H_m, d \mu_m) \;. \label{H1L2}
\end{equation}
Having constructed the 1-particle space, we turn to the $n$-particle space, defined as the symmetric tensor product
\begin{equation} 
{\cal H}_n = S_n {\cal H}_1^{\otimes n}  \;, \label{hn}
\end{equation}
where $S_n$ stands for the permutation operator 
\begin{equation} 
S_n {\hat f}_n = \frac{1}{n!} \sum_{P} {\hat f}_n(p_{P_1}, ...,p_{P_n}) 
\end{equation}
and the sum runs over all permutations of the momenta. The $n$-particle space ${\cal H}_n$ is spanned by the vectors
\begin{equation} 
\frac{1}{\sqrt{n!}} \Pi_{i=1}^{n} \;a^{\dagger}_{h_i} |0\rangle   \;. \label{np}
\end{equation}
which turn out to be isomorphic \cite{Scharf:1996zi} to the symmetric tensor product of  test functions $S_n {\hat h}_1\otimes ....... \otimes {\hat h}_n$. The appearance of the symmetric tensor product in eq.\eqref{hn} is a consequence of the commutation relations fulfilled by the creation and annihilation operators. In particular, as the creation operators commute among themselves, their product automatically projects into the symmetrized product of the corresponding test functions. \\\\The Hilbert space ${\cal H}$ of the free field is  obtained by taking the direct sum \cite{Scharf:1996zi,Guido:2008jk}
\begin{equation} 
{\cal H} = \bigoplus_{n=0}^{\infty} {\cal H}_n \;, \label{hf}
\end{equation}  
where ${\cal H}_0$ refers to the vacuum state $|0\rangle$. This procedure is known as the Fock space construction.

\subsubsection{Weyl operators}\label{Weyl}

It should be noted that the operators $(a_h, a^\dagger_h)$ are unbounded operators \cite{BR}, a well known feature stemming from the canonical commutation relations. As the construction of the Bell inequality in Quantum Field Theory requires the use of bounded operators, it is helpful to replace the unbounded object $\varphi(h)$ by its bounded version, a task achieved by introducing the Weyl operators, namely 
\begin{equation} 
\varphi(h) \rightarrow W_h = e^{i {\varphi}(h) }\;, \label{Weyl}
\end{equation}
The Weyl operators are unitary operators:
\begin{equation} 
 W_h  W_h^{\dagger} = W_h^{\dagger}  W_h =1 \;, \qquad W_h^{\dagger}= W_{-h} = e^{-i {\varphi}(h) }\;. \label{Weyl1}
\end{equation}
Making  use of  the relation
\begin{equation}
e^A \; e^B = \; e^{ A+B +\frac{1}{2}[A,B]}, \label{exp_AB}
\end{equation} 
valid for two operators $(A,B)$ commuting with $[A,B]$, it turns out that the canonical commutation relations can be recast in the form 
\begin{equation}
W_h \;W_{h'}  =   e^{- \frac{1}{2} [{\varphi}(h), {\varphi}(h')] }\;W_{(h+h')} = e^{ - \frac{i}{2} \Delta_{\textrm{PJ}}(h,h')}\;W_{(h+h')} \;,   \label{algebra} 
\end{equation} 
where $\Delta_{\textrm{PJ}}(h,h')$ is the smeared Pauli-Jordan expression~\eqref{pauli_jordan_smeared}. Moreover, setting 
$\varphi(h) = (a_h + a_h^\dagger)$ and using the canonical commutation relations~\eqref{ccrfg}  and \eqref{exp_AB}, one evaluates  the vacuum expectation value of the Weyl operator $W_h$, finding 
\begin{equation} 
\langle 0 \vert  W_h  \vert 0 \rangle = \; e^{-\frac{1}{2} {\lVert {h}\rVert}^2}, \label{vA}
\end{equation} 
where ${\lVert { h}\rVert}^2 \equiv \langle { h} | { h} \rangle = \int d\mu(p)_m |{ h}(p)|^2$.  In particular, if $supp_f$ and $supp_g$ are space-like separated, causality ensures that the Pauli-Jordan function vanishes. From the above properties, it follows the  relation 
\begin{equation} 
\langle 0 \vert W_f W_{g}  \vert 0 \rangle =  \langle 0 \vert W_{({ f} + { g})} \vert 0 \rangle =
\; e^{-\frac{1}{2} {\lVert { f}+{ g} \rVert}^2}. \label{vAhh}
\end{equation}
The  Hilbert space ${\cal H}$ can be re-expressed in terms of the Weyl operators  \cite{Guido:2008jk} as
\begin{equation} 
{\cal H} = {\rm span}  \{ W_{h_1}...W_{h_n} |0\rangle, \; (h_1,...,h_n) \in{\cal S}( \mathbb{R}^4) \} \;, \label{Hw}
\end{equation}
meaning that the states of ${\cal H}$ can be obtained by taking suitable products and combinations of the Weyl operators. \\\\ {\underline {\bf Comment}} As we shall see in Sect.\eqref{Vn} and in Sect.\eqref{HK}, expression \eqref{Hw} relies on two key structural features of Quantum Field Theory. The first one is that the set of Weyl operators $\{ e^{i\varphi(h)}, supp(h) \subseteq {\cal O} \}$ gives rise to a von Neumann algebra ${\cal A}({\cal O})$  \cite{Summers:1987fn,Summ,Summers:1987ze}, where ${\cal O}$ stands for a generic  open   region of the Minkowski spacetime. The second one  is that, due to the Reeh-Schlieder Theorem, the vacuum state $|0\rangle$ is cyclic and separating for ${\cal A}({\cal O})$. As a consequence, any state of ${\cal H}$ can be arbitrarily well approximated by acting on the vacuum $|0\rangle$ with a suitable set of elements of ${\cal A}({\cal O})$. \\\\Let us conclude  by remarking that the construction outlined above  can be performed  in Quantum Mechanics as well  \cite{BR,SF}. Here, the Weyl operators are the so-called unitary displacement operators and the corresponding states  are nothing but  the coherent states.

\section{Von Neumann algebras and their classification}\label{Vn}

We provide here a  concise summary on some basic  notions on von Neumann algebras and on the Tomita-Takesaki construction. The material is by no means to be considered exhaustive. We refer to the classical book by \cite{BR} for more details and proofs of the statements. 

\subsection{Algebras, $^*$-algebras, $C^*$-algebras}

Let ${\cal F}$ be a vector space over $\mathbb{C}$. ${\cal F}$ is called an algebra if it is equipped with a multiplication law which associates the product $AB$ to each pair $(A,B) \in {\cal F}$. The product is assumed to be associative and distributive
\begin{eqnarray} 
A (BC) & =& (AB)C \;, \nonumber \\
A(B+C) & = & AB + AC \; \nonumber \\
\alpha \beta (AB) & =& (\alpha A ) (\beta B) \;, \qquad (A,B) \in {\cal F}\;, (\alpha, \beta) \in \mathbb{C} \;. \label{ass}
\end{eqnarray} 
The algebra ${\cal F}$ is called commutative or Abelian, if the product is commutative: $(AB) = (BA)$.\\\\\underline{\bf Involution} A mapping $A\in {\cal F}$ $\rightarrow$ $A^{\dagger}\in {\cal F}$ is called an involution, or adjoint operation, if it has the following properties: 
\begin{eqnarray} 
(A^{\dagger})^{\dagger} & = & A \;, \nonumber \\
(AB)^{\dagger} & = & B^{\dagger} A^{\dagger} \;, \nonumber \\
(\alpha A + \beta B)^{\dagger} & = & {\alpha}^*  A^{\dagger} + {\beta}^* B^{\dagger}   \;, \label{inv} 
\end{eqnarray} 
where ${\alpha}^*, {\beta}^*$ are the complex conjugates of $\alpha, \beta$. \\\\\underline{\bf Normed Algebras}  The algebra $\cal F$ is a normed algebra if a norm $||\;\;||: {\cal F} \rightarrow \mathbb{R}$ is assigned: 
\begin{eqnarray} 
|| A|| &  \ge & 0 \;, \qquad ||A||= 0 \Rightarrow A=0 \;, \nonumber \\
|| A + B|| & \le & ||A|| + ||B|| \;, \nonumber \\
|| A B|| & \le & ||A|| ||B||  \;. \label{normalg}
\end{eqnarray} 
\underline{\bf Banach Algebra} If the normed algebra ${\cal F}$ is complete with respect to the norm $||\;\;||$, {\it i.e.} if every Cauchy sequence converges to an element of $\cal F$, ${\cal F} $ is called a Banach algebra. \\\\\underline{\bf Banach *-algebra}  A normed complete algebra equipped with an involution $\dagger$ and such that 
\begin{equation} 
|| A^{\dagger}|| = ||A|| \;, \label{stB}
\end{equation}
is called a Banach * -algebra. \\\\\underline{\bf $C^*$-algebra} A $C^*$-algebra is a Banach *-algebra with the property 
\begin{equation} 
|| A^{\dagger} A || = ||A||^2  \qquad \forall A \in {\cal F} \;. \label{starcond}
\end{equation}
This property is called the *-condition. The *-condition automatically implies that $||A^{\dagger}|| = ||A||$.

\subsection{Convergence in ${\cal B}({\cal H})$} 
Let ${\cal H}$ be a Hilbert space and let ${\cal B}({\cal H})$ denote the set of all bounded operators acting on 
${\cal H}$. Several topologies can be defined on ${\cal B}({\cal H})$, namely 
\begin{itemize} 
\item \underline{ Norm topology.} It is defined by the norm operator 
\begin{equation} 
|| {\cal A} || = sup \left\{  \frac{||{\cal A}\psi||}{||\psi||} \;; \forall \psi \neq 0\;, \psi \in {\cal H} \right\} \;. \label{normA}
\end{equation}
A sequence of operators $(A_i)_{i \in \mathbb{N}} \in {\cal B}({\cal H})$  converges in the norm topology to $A\in {\cal B}({\cal H})$ if $||A_i -A||$ converges to zero 
\begin{equation} 
\textrm{ lim}_{i \rightarrow \infty}  ||A_i -A|| = 0 \;. \label{normconv} 
\end{equation} 
\item \underline{Strong topology.} It is defined by the semi-norm $\nu_\psi(A)$ 
\begin{equation}
\nu_\psi(A)  = ||A \psi || \;, \qquad \psi\in {\cal H} \label{strt} 
\end{equation}
A sequence of operators $(A_i)_{i \in \mathbb{N}} \in {\cal B}({\cal H})$  converges in the strong  topology to $A\in {\cal B}({\cal H})$ if $||A_i \psi -A\psi ||$ converges to zero 
\begin{equation} 
\textrm{ lim}_{i \rightarrow \infty}  ||A_i \psi -A\psi || = 0 \;, \forall \psi \in {\cal H}  \;. \label{strnormconv} 
\end{equation} 
\item \underline{Weak  topology.} It is defined by the semi-norm $\nu_{\psi\varphi}(A)$ 
\begin{equation}
\nu_{\psi\varphi}(A)  = \langle \varphi \; | \; A \psi \rangle \;, \qquad \varphi, \psi \in {\cal H} \label{wtrt} 
\end{equation}
A sequence of operators $(A_i)_{i \in \mathbb{N}} \in {\cal B}({\cal H})$  converges in the weak  topology to $A\in {\cal B}({\cal H})$ if 
\begin{equation} 
\textrm{ lim}_{i \rightarrow \infty}  | \langle \varphi \; | \; A_i \psi -A\psi \rangle | = 0 \;, \forall \varphi, \psi \in {\cal H}  \;. \label{wtrnormconv} 
\end{equation} 
\item From the Cauchy-Schwarz inequality and from the properties of the norm operator, it follows that the convergence given by the norm topology is the strongest one. Norm convergence implies strong convergence which, in turn, implies weak convergence. \\\\The pair $\left( {\cal B}({\cal H}), ||\;\;||\right) $ where $||\; \;||$ is the norm operator, eq.\eqref{normA}, is a $C^*$-algebra. 
\end{itemize}

\subsection{Von Neumann algebras} 

\underline{\bf The commutant} For any subset ${\cal F} \subset {\cal B}({\cal H}) $, the commutant ${\cal F'}$ is the set of all operators of ${\cal B}({\cal H})$ which commute with each element of ${\cal F}$ 
\begin{equation} 
{\cal F'} = \{ B \in {\cal B}({\cal H}) \; \textrm{ such that} \; AB=BA\;, \forall A\in {\cal F} \}  \;. \label{comm}
\end{equation} 
Von Neumann algebras can be defined in two equivalent ways.\\\\\underline{\bf Definition} A *-subalgebra ${\cal F} \in {\cal B}({\cal H})$ containing the identity is a von Neumann algebra if it is closed under weak topology. \\\\\underline{\bf Definition} A *-subalgebra ${\cal F} \in {\cal B}({\cal H})$ containing the identity is a Von-Neumann algebra if ${\cal F}$ is such that
\begin{equation} 
{\cal F} = ({\cal F'})' \;. \label{MMp}
\end{equation}
The equivalence between these two definitions is ensured by von Neumann's bicommutant Theorem:\\\\\underline{\bf Bicommutant Theorem}  Let   ${\cal F}$ be a *-subalgebra of  ${\cal B}({\cal H}) $. The following conditions are equivalent
\begin{eqnarray} 
i) \;\;\;& {\cal F} & =  ({\cal F'})'  \;, \nonumber \\
ii) \;\;\;& {\cal F} & \;\;\textrm{is closed under weak topology} \;, \nonumber \\
iii) \;\;\;& {\cal F} &\;\; \textrm{is closed under strong topology} \;. \label{bicomm} 
\end{eqnarray} 
\underline{\bf Definition} A vector $ \Omega \in {\cal H}$ is a cyclic vector for the von Neumann algebra ${\cal F}$ if 
\begin{equation} 
 \textrm {span} \{ {\cal F} \Omega\} \; \textrm{is dense in} \;{\cal H}. \;. \label{cycl}
\end{equation} 
\underline{\bf Definition} A vector $ \Omega \in {\cal H}$ is called separating for a von Neumann algebra ${\cal F}$ if the condition $A \Omega=0$, $A \in {\cal F}$ implies that $A=0$. \\\\\underline{\bf Remark} Let ${\cal F}$ be a von Neumann algebra and $\Omega \in {\cal H}$. The following statements are equivalent:
\begin{eqnarray} 
i) \;  \Omega &\; & \textrm{is cyclic for} \; {\cal F} \;, \nonumber \\
ii) \; \Omega &\; & \textrm{is separating  for} \; {\cal F'} \;.  \label{mmm}
\end{eqnarray}

\subsection{States and GNS construction} 

From the abstract algebra of operators one can  define states as linear maps, or linear forms, taking operators into complex numbers. A linear map $\phi: {\cal F} \rightarrow \mathbb{C}$ is defined such that 
\begin{eqnarray}
	\phi(aA + bB) = a\phi(A)+ b\phi(B) 
\end{eqnarray} 
where $A, B \in {\cal F}$ and $a, b \in \mathbb{C}$. The linear map is bounded if  
\begin{eqnarray}
	|\phi(A)| \le  c ||A||
\end{eqnarray} 
where $c$ is a positive number. The lowest bound for $c$ defines the norm of $\phi$ 
\begin{eqnarray}
	||\phi|| \equiv  \sup_{A\in {\cal A}}\frac{|\phi(A)|}{||A||}
\end{eqnarray} 
The linear map is said to be real if $\phi(A^{\dagger}) = \phi(A)^{\ast}$. It is said to be positive if $\phi(A^{\dagger} A) \ge 0$ and it is said to be normalized if $||\phi|| = 1$. \\\\A state is defined as a normalized, positive linear map.  In the Quantum Theory interpretation, a state acting on an operator gives the expectation value, or mean value, of the observable related to the operator. \\\\A very important result is known as the Gelfand-Naimark-Segal (GNS) construction: each state $\omega$ acting on a von Neumann algebra ${\cal F}$ defines a Hilbert space ${\cal H}_{\omega}$ and a representation $\pi_{\omega}$ of ${\cal F}$ given by linear operators acting on ${\cal H}_{\omega}$.  This gives a different point of view of the whole mathematical construction of Quantum Mechanics. In this case, the algebra of observables and the linear maps are the fundamental entities. The Hilbert space, along with a representation, arises from this underlying structure. \\\\The GNS construction proceeds as follows: 
\begin{itemize}
	\item  Noting that the algebra ${\cal F}$ itself is a linear space over the complex $\mathbb{C}$, to make it into a Hilbert space one must provide a definition of the inner product. Given a state $\omega$, one can naturally define an inner product as:
	\begin{eqnarray}
		\langle A | B \rangle \equiv \omega(A^{\dagger}B),\;\; \forall \; A, B \in {\cal F} \label{gnsinn}
	\end{eqnarray} 
	By the properties of $\omega$, expression \eqref{gnsinn} is positive semi-definite $\langle A | A \rangle  \ge 0$ and satisfies the Cauchy-Schwarz inequality: $|\langle A | B \rangle|  \le \langle A | A\rangle \langle B | B \rangle $
	
	\item The inner product has an ambiguity related to the subset ${\cal J}$ of operators $X$ that satisfy $\omega(X^{\dagger} X) = 0$. This has the following consequence: if $X \in {\cal J} \subset {\cal F}$, then for all $A \in {\cal F}$ it follows that $AX \in {\cal J}$, as it stems  from the Cauchy-Schwarz inequality:  $\omega(X^{\dagger}A^{\dagger}AX)   \le \omega(A^{\dagger}AXX^{\dagger}A^{\dagger}A) \omega(X^{\dagger} X) = 0$. The set ${\cal J}$ is called left ideal since it is invariant under multiplication by any other operator from the left. This set depends on the state and in this context is known as the Gelfand left ideal of $\omega$. One then defines the equivalence class of operators ${\cal F}/{\cal J}$ by $[A] = \left\{A + X,\;\; A \in {\cal F},\;\; \forall X \in {\cal J}\right\}$. It is clear that $\omega$ depends only from the class $\omega ([A]^{\dagger} [B]) = \omega(A^{\dagger}B)$.\\\\The space ${\cal H}_{\omega}$ is thus defined as the space  ${\cal F}/{\cal J}$ completed with limits under the norm just defined. That is, a state $\Psi$ of ${\cal H}_{\omega}$ is defined by an element $[A]$ of ${\cal F}/{\cal J}$. 
	
	\item The product  between elements of ${\cal F}$ naturally defines the action of operators on ${\cal H}_{\omega}$. It associates to  each element $A \in {\cal F}$ an operator $\pi_{\omega}(A)$ acting on states of ${\cal H}_{\omega}$ in an obvious way:
	\begin{eqnarray}
		\pi_{\omega}(A) | B \rangle \equiv [A][B] = [AB] 
	\end{eqnarray} 
	The set of operators $\pi_{\omega}(A)$ forms the algebra ${\cal B}({\cal H}_\omega)$ of bounded operators acting on ${\cal H}_\omega$.
	
	\item A cyclic state $|\Omega \rangle$ in a Hilbert space  ${\cal H}_{\omega}$ is one such that $\pi_{\omega}(A)|\Omega \rangle$ is dense in  ${\cal H}_{\omega}$, that is, any other state in ${\cal H}_{\omega}$ can be constructed arbitrarily well from $|\Omega\rangle$ by acting with the operators of ${\cal B}({\cal H}_\omega)$. In a von Neumann algebra there is a natural cyclic vector state defined by the unit of the algebra
	\begin{eqnarray}
		|\Omega \rangle \equiv [\mathbf{1}]
	\end{eqnarray} 
	It follows that 
	\begin{eqnarray}
		\omega(A) = \omega( \mathbf{1} A  \mathbf{1})= \langle \Omega | A \Omega \rangle =  \langle \Omega | \pi_{\omega}(A) |\Omega \rangle 
	\end{eqnarray} 
	which is the usual form of the expectation value. In fact one can write expectation values for any vector $\Psi \in {\cal H}_{\omega}$. Any such vector defines a state
	\begin{eqnarray}
		\omega_{\Psi}(A)  =  \langle \Psi | \pi_{\omega}(A) |\Psi \rangle =  \langle \Omega |\pi_{\omega}(B)^{\dagger} \pi_{\omega}(A)\pi_{\omega}(B) |\Omega \rangle  = \omega(B^{\dagger} A B)
	\end{eqnarray} 
	where it was used that there exists $B$ such that $\pi_{\omega}(B) |\Omega \rangle$ can be made arbitrarily equal to $|\Psi\rangle $, because $|\Omega \rangle$ is cyclic. \\\\The vectors such as $|\Psi\rangle $ are called the vector states of the representation $\pi_{\omega}$. There are also more general states that cannot be written as vectors under the representation. More generally, the set of states of the form
	\begin{eqnarray}
		\omega_{\rho} (A) = Tr \left( \rho \pi_{\omega}(A)\right)   
	\end{eqnarray} 
	where $\rho$ is a positive trace class element of the algebra ${\cal B}({\cal H}_\omega)$,  are called normal states of the von Neumann algebra. 
	
	\end{itemize}

\subsection{Modular structure and the Tomita-Takesaki Theorem} \label{TTT}

The modular theory of Tomita-Takesaki \cite{TT} is a fundamental tool in the study of the von Neumann algebras, with very relevant applications in Quantum Field Theory and Statistical Mechanics \cite{Borchers:2000pv}. \\\\ 
\underline{\bf Definition of the anti-linear operators $S$ and $S^\dagger$} 
Let $\Omega \in {\cal H}$ be a separating and cyclic vector for the Von Neumann algebra $\cal F$, then $\Omega$ is also separating and cyclic for the commutant ${\cal F'}$. Define the operators \cite{Summers:2003tf,Guido:2008jk}

\begin{eqnarray} 
& \bullet &\;\; S\; A\; \Omega = A^{\dagger} \; \Omega  \;, \nonumber \\
& \bullet &\;\; S^{\dagger}\; A'\; \Omega = (A')^{\dagger} \; \Omega \;. \label{SSd}
\end{eqnarray} 
The operators $(S,S^{\dagger})$ are anti-linear and 
\begin{equation} 
SS = S^{\dagger} S^{\dagger} = 1 \qquad S \Omega = S^{\dagger} \Omega = \Omega\;. \label{SSD1}
\end{equation} 
\underline{\bf Polar decomposition} 
\begin{equation} 
S = J \Delta^{1/2} \;, \label{PD}
\end{equation} 
where $J$ is an anti-linear operator, called the conjugation operator. $\Delta$ is a self-adjoint, positive operator called the modular operator. The following properties hold \cite{Summers:2003tf,Guido:2008jk}

\begin{eqnarray} 
J \Delta^{1/2} J & = & \Delta^{-1/2}  \;, \nonumber \\
\Delta & = & S^{\dagger} S \;, \qquad
\Delta^{-1}  =  S S^{\dagger} \;, \nonumber \\
J^{\dagger} & = & J \;, \qquad J^2 = 1 \;. \label{DJprop} 
\end{eqnarray} 
\underline{\bf Tomita-Takesaki Theorem} 

\noindent Let ${\cal F}$ be a von Neumann algebra with a cyclic and separating vector $\Omega \in {\cal H}$. Let $\Delta$ be the associated modular operator and $J$ the conjugation operator. Then 
\begin{eqnarray} 
& \bullet & \;\;\; J {\cal F}J = {\cal F'} \;, \qquad J {\cal F'} J = {\cal F}  \;, \nonumber \\
& \bullet & \;\;\; \textrm{there exists a one parameter family of unitary operators} \nonumber \\ & & \Delta^{it}, \; t \in \mathbb{R},  \textrm{such that}\;\;   \Delta^{it} {\cal F} \Delta^{-it} = {\cal F} \;. \label{TTtheorem}
\end{eqnarray} 
From this Theorem one sees that the modular conjugation $J$ maps the algebra ${\cal F}$ into its commutant ${\cal F}'$. As we shall see, this result has far reaching applications in the Quantum Field Theory formulation of the Bell inequality. Also, the unitary operator $\Delta^{it} = e^{it \log(\Delta)}$ defines the modular automorphism group which leaves ${\cal F}$ invariant.

\subsection{Spectrum of the modular operator and Connes classification of the Von Neumann algebras} 

There are different types of von Neumann algebras \footnote{More precisely, we consider only factors. These are von Neumann algebras ${\cal A}$ such that ${\cal A} \cap {\cal A}'$ contains only multiples of the identity. In other words, the center is trivial. This is no limitation because every von Neumann algebra can be decomposed into factors}.  In order to understand  their classification, it is instructive to consider the algebraic structure of certain special operators in the algebra called projectors\footnote{The use of projectors to classify the types of algebras is due to  von Neumann and Murray \cite{mvn1}, see also chapter $1$ of \cite{sunder} and \cite{Sorce:2023fdx} for a  pedagogical recent review}. By definition, a projector is an operator $P \in {\cal B}({\cal H}) $ that satisfies $P^2 = P= P^{\dagger}$. It defines a subspace $P{\cal H} \subset {\cal H}$.  This inspires a notion that a projector $P$ is larger or smaller than another projector $Q$ by comparing the cardinality of  $P{\cal H}$ and $Q{\cal H}$. Let's consider now a set of definitions that allows for a more precise and purely algebraic characterization of this notion.

\begin{itemize}
	
	\item  The algebraic characterization of this ``size relation'' is achieved by defining that two projections $P, Q \in {\cal A}$ in a generic algebra ${\cal A}$ satisfy
	\begin{eqnarray}
		Q\le P \;\; \text{if} \;\; QP=Q
	\end{eqnarray}
	It is clear that the smallest projection is $0$ and the largest is the identity $\mathds{1}$.

	\item  It also follows a notion of equivalence. One says that two projectors $P$ and $Q$ are equivalent, $P \sim Q$,  if there is a partial isometry $V \in {\cal A}$ such that  $V^{\dagger}V = Q$ and $VV^{\dagger} = P$. To understand this definition note that, in the case ${\cal A} =  {\cal B}({\cal H})$, $\psi \in Q{\cal H} \subset {\cal H}$ is mapped under $V$ to $V\psi = V V^{\dagger} V \phi = P V\phi \in P{\cal H} \subset {\cal H}$. That is, every vector in the space  $Q{\cal H} $ can be mapped to a vector in  $P{\cal H}$ by the partial isometry $V$. One can intuitively say that the spaces $P{\cal H}$ and $Q{\cal H}$ are of the ``same size'', in the sense that the partial isometry relates the range of $Q$ to the range of $P$ . \\\\Having a notion of size, it is possible then to construct the following definition of finite and infinite projectors.
	\item  A projector $P$ is said to be finite if it is not equivalent to a smaller projector. Conversely, a projector $P$ is infinite if it is equivalent to a smaller projector $Q$, that is
	\begin{eqnarray}\label{infproj}
		P \;\; \text{is infinite} \Rightarrow Q <  P \;\; \text{and} \;\; Q\sim P
	\end{eqnarray}
	Intuitively, one can try to ``reduce the size'' of  $P$, but if it is infinite it will remain infinite.  An important observation is that, since $\mathds{1}$ is the largest projector, if there is at least one infinite projector then $\mathds{1}$ is certainly infinite. On the other hand, if $\mathds{1}$ is finite then all projectors are finite.
	\begin{eqnarray}\label{infprojunit}
		P \;\; \text{is infinite} &\Rightarrow& \mathds{1}\;\;\text{is infinite}\nonumber\\
		\mathds{1} \;\; \text{is finite} &\Rightarrow& P \;\;\text{is finite}
	\end{eqnarray}
	Thus, to check if there is any infinite projector one just needs to check the identity.
	
	\item   One of the most important results in this subject is that all infinite projectors are equivalent. Or, which is the same, all infinite projectors are equivalent to the identity. 
	\begin{eqnarray}\label{infprojunitequiv}
		P \;\; \text{is infinite} &\Rightarrow& P \sim \mathds{1}
	\end{eqnarray}
	
\end{itemize}

These properties of projectors can be used to characterize the types of algebras
\begin{itemize}
	\item A type $I$ algebra contains a nonzero minimal projector. A minimal projector is necessarily a finite projector such that there are no nonzero projector smaller than it. This is the same as the algebra admitting an irreducible representation.
	
	\item A type $II$ algebra contains a nonzero finite projector, but no nonzero minimal projectors.  This translates to the fact that such algebras don't have  irreducible representations.
	
	\item A type $III$ algebra contains no nonzero finite projectors. All projectors are infinite and equivalent to the identity. 
\end{itemize}

\begin{itemize}
	
	\item A type $I$ factor has the standard form ${\cal B}({\cal H}) \otimes \mathds{1}_{P{\cal H}}$, where $P$ is the minimal projector. If the dimension of ${\cal H}$ is $n$ the algebra is said to be of type $I_n$, $n$ can be infinite. This is the usual algebra of operators one encounters in non-relativistic Quantum Mechanics. 
	 
	\item If a type $II$ factor only contains finite projections it is called type $II_1$, but it differs from type $I$ in the sense that there are no minimal projections. If there is an infinite projector in a type $II$ it is said to be of type $II_{\infty}$. The standard form of a type $II_{\infty}$  factor ${\cal A}$ is  ${\cal B}({\cal H}) \otimes P{\cal A}P$, where $P \in {\cal A}$ is  a finite projector. The algebra $P{\cal A}P$ is a von Neumann algebra acting on  $P{\cal H}$. It contains no minimal projector, since otherwise that would imply that ${\cal A}$ had a minimal projector. The identity element in $ P{\cal A}P$ is $P$, which is finite, therefore this is a finite algebra without minimal projectors and thus it is of type $II_1$. So, a type  $II_{\infty}$ factor has the standard form of the tensor product of bounded operators and a type $II_1$ factor.
	
	\item The classification of the different type $III$ factors is more complicated and was constructed by Connes \cite{Connes}. The construction heavily uses modular theory. A key theorem is the following: consider a modular operator $\Delta_{\Omega}$, associated with a cyclic and separating vector $\Omega$ for a factor ${\cal A}$, then, if  ${\cal A}$ is type $III$ the spectrum of  $\Delta_{\Omega}$ must include zero. \\\\This is related to the fact that in a type $III$ algebra the notion of a finite nonzero trace don't exist and one needs to consider a more general notion of states. An important property of projectors is that every positive operator (a candidate for a density matrix) can be approximated arbitrarily well by a positive linear combination of projections. Thus, one can rephrase the classification of algebras in terms of states: an algebra is type $I$ if there are pure states (minimal projector) with respect to it. If there are only mixed states (nonminimal finite projectors) the algebra is type $II$. But if every projector is infinite, then the density matrix cannot be normalized and therefore, in a type $III$ algebra there are no normalizable tracial states. But one can still define a notion that assigns expectation values to operators, except that they are not always finite. \\\\This more general notion of states relies on a linear function $\phi : {\cal A}_{+} \rightarrow [0, \infty]$ that maps positive operators to positive numbers, possibly infinite. It is faithfull, meaning that $\phi(A) = 0 \Leftrightarrow A =0$. It is normal, meaning that given a family of positive operators $A_{\alpha} \in{\cal A}_{+}$, $\phi(sup_{\alpha} A_{\alpha}) = sup_{\alpha} \phi(A_{\alpha})$ and is semifinite in the sense that the set of operators ${\cal A}_{\phi, +} = \left\{A, \phi(A) < \infty   \right\} \subset {\cal A}_{+} $ is  dense in ${\cal A}_{+}$. The map $\phi$  is called a faithfull normal semifinite weight and it generalizes the notion of state (or projection) for the case of infinite von Neumann algebras. Every von Neumann algebra admits a faithful normal semifinite weight (proposition 2.7.13 in \cite{BR}). Connes used this notion to provide a finer classification of the type $III$ algebras. The idea is to single out a class of operators ${\cal A}_{\phi}$ that are finite under $\phi$, in the sense that $\phi(A^{\dagger} A) < \infty$ and thus define an inner product in ${\cal A}_{\phi}$ making it into a Hilbert space using the GNS construction. The algebra  ${\cal A}$ acts on this Hilbert space and one can follow the usual steps and construct a modular operator $\Delta_{\phi}$ associated with the weight $\phi$. Connes then showed that the spectrum of $\Delta_{\phi}$, for all $\phi$, can be used to classify the type $III$ algebras as, see \cite{Summers:2003tf} for the notations:
	\begin{itemize}
		\item  $ {\cal A}$ is type $III_0$ if $ \sigma_{sp}( {\cal A}) = \{0, 1 \}$
		
		\item  $ {\cal A}$ is type $III_{\lambda}$ if  $\sigma_{sp}( {\cal A}) = \{ 0 \} \cup \{\lambda^n; \;\; n\in \mathds{Z}$ for some $0 < \lambda < 1 \}$
		
		\item  $ {\cal A}$ is type $III_{1}$ if  $\sigma_{sp}( {\cal A}) = [0, \infty)$ 
	\end{itemize}	 
	where $\sigma_{sp}( {\cal A}) = \bigcap_{\phi}\left\{\text{spectrum}\; \Delta_{\phi}; \;\;  \phi \;\text{a faithful normal semifinite weight} \right\}$ is the intersetion of the spectrum of all the modular operators associated with the faithful normal semifinite weights.

\end{itemize}

\section{Elements of Algebraic Quantum Field Theory} \label{HK}

\subsection{Nets of local algebras} 

One of the basic formalisms of the Algebraic Quantum Field Theory regards nets of $C^*$-algebras  and it yields a framework to deal with fields in a region of spacetime and their properties when taking relativity into account. More specifically, a net of algebras is a mapping:
\begin{equation}
O \rightarrow  {\cal A}(O) \;, \label{nnn}
\end{equation} 
where $O$ is a region of spacetime\footnote{The region $O$ is usually an open bounded subset in Minkowski space as given, for instance,  by double cones. A double cone is the non-empty intersection of the causal future cone of a point $x$ with the causal past cone of a point $y$.} and ${\cal A}(O)$ is a von Neumann algebra or a $C^*$-algebra of  operators localized  in $O$. The  self-adjoint elements of the algebra are identified with the local observables. \\\\Those algebras satisfy certain intuitive properties such as isotony and locality/causality:

\begin{itemize}
\item
Isotony: If $O_1 \subseteq O_2$ , then  there is an embedding ${\cal A}(O_1)  \rightarrow {\cal A}(O_2)$. This is also called an inductive system.
\item
Locality/Causality: If $O_1$, and  $O_2$ are spacelike separated, the algebras commute:
$[{\cal A}(O_1), {\cal A}(O_2)] = 0.$
\end{itemize}
One motivation for isotony is the assumption that an observable measurable in a region $O_1$ is thus also measurable in any region $O_2$ for which $O_1$ is a subset. Also it means that there is a $C^*$-algebra that may be defined as the inductive limit of the local algebras ${\cal A}$. As for the Locality/Causality, it is a way to take into account the spacetime features of special relativity.

\subsection{The Haag-Kastler axioms} 

Together with the above isotony and locality/causality properties more general features may be assumed and they provide an axiomatic framework from which it is possible to build an Algebraic Quantum Field Theory. Those constitute the so called Haag-Kastler axioms \cite{Guido:2008jk, Haag:1992hx}. Here we restate them using the notion of causal complement, that is, if $O$ is a set of points in Minkowski space, its causal complement $O'$ is the set of points that are space-like separated from all points of $O$. We call $(O')' \equiv O''$ the completion of $O$. When $O = O''$ it is causally complete. Given a net $O \rightarrow  {\cal A}(O)$ for which $O = O''$, one has:
\begin{itemize}
\item Isotony: If $O_1 \subseteq O_2$ , then  ${\cal A}(O_1) \subseteq {\cal A}(O_2)$.
\item Locality: If $O_1 \subseteq O_2'$, then  ${\cal A}(O_1) \subseteq {\cal A}(O_2)'$
\item (Haag) Duality: ${\cal A}(O') = {\cal A}(O)'$
\item Poincaré symmetry: The action of the Poincaré group on the net is an automorphism  $\mathcal{A}_g (O) = {\cal A}(g O)$, where $g$ is an element of the proper orthochronous Poincaré group.

\end{itemize}
Given a free scalar field with a continuous unitary representation $U(\Lambda, a)$ of the Poincaré group acting on the Hilbert space $\cal H$, the action of this group on the net of algebras is covariant, in the sense that for a spacetime region $O$, the transformed algebra satisfies:
\begin{align}
\label{FFP+}
     U(\Lambda, a) {\cal A} (O) U(\Lambda, a)^{-1} = {\cal A}(\Lambda O + a)
\end{align}
where $\Lambda$ is a proper orthochronous Lorentz transformation and 
$a$ is a translation.

\subsection{The Reeh-Schlieder Theorem} 

We have  seen that for a free scalar field the relation given by eq.\ref{FFP+} holds. It is customary to supplement, the Haag-Kastler axioms, for the net of local algebras of the free scalar field, by the following: \\\\reminding that a  representation $\pi$ of a net  on a Hilbert space ${\cal H}$ is a family $\{\pi_O\}$,  where $\pi_O$ is a representation of ${\cal A}(O)$ on ${\cal H}$  such that, if
 ${\cal A}(O_1) \subseteq {\cal A}(O_2)$ , the restriction of  $\pi_{O_2}$ to ${\cal A}(O_1)$ is  $\pi_{O_1}$. 
 
\begin{itemize}
\item Positive energy: the   spectrum of the generators of the translation subgroup lies in the  forward light cone.
\item Vacuum: there exists a unique translation invariant vector $\Omega$\footnote{Aside a multiplicative constant.}. The set $\{\pi_O(R)\Omega\},$ $R \in {\cal A}(O)$,  where $O$ is a double cone, is dense in ${\cal H}$.
\item Additivity: For the net $O \rightarrow  {\cal A}(O),$  it holds that $ O = \bigcup_i O_i  \implies  {\cal A}(O) =  \lor_i {\cal A}(O_i)$
\end{itemize}
Endowed with a  net $O \rightarrow {\cal  A}(O),$  that satisfies the above assumptions one has that, {\it for all double cones $O,$ the vacuum $\Omega$ is cyclic for ${\cal A}(O)$, as well as separating for every local algebra}. This property is usually called the Reeh-Schlieder Theorem, and it is one of the more striking results in AQFT. Essentially the theorem states that acting on the vacuum  $\Omega$ with appropriate elements of ${\cal A}(O)$ it is possible to approximate as closely as desired any vector of the Hilbert space.  One question that arises is that the Theorem allows that from some  $O$ we could approximate a state in a spacelike separated region $O'$, that could raise questions about the violation of locality. The point that clarifies this is that the vacuum is a highly correlated state, and the theorem is made possible by means of those correlations between operators in the two spacelike separated  regions \cite{Witten:2018zxz}.

\subsection{The Bisognano-Wichmann results}\label{bwth}

Let us remind here   that we consider a free field $\varphi$ as a map  that takes elements  $f \in {\cal S}({\mathbb{R}}^4)$ of the Schwartz space to $\varphi(f)$, the self-adjoint operators that generate the one parameter group through Weyl unitary operators $W_{\lambda \hat{f}}\; $, see Section \ref{fock1}). It is customary to call the next properties (1 to 5)  as the G\aa rding- Wightman axioms, see  \cite{Guido:2008jk}:

\begin{description}

  \item[1] The map $f \to \varphi(f)$ is an operator valued distribution.
  \item[2] For all of such maps (fields)  there is a dense common invariant domain $D$, and the set of $\varphi(\bar{f}) $, $\bar{f}$ being the complex conjugate, is contained in the set of  $\varphi(f)^{\dagger}\;.$
  \item[3] There is a strongly continuous, positive energy unitary representation\footnote{When for every element $h$ of the Hilbert space and for $g \to g_0$ one has $U(g)h \to U(g_0)h$ }  $U$ of the proper orthochronous Poincaré group satisfying $U (g)\phi(f)U (g)^{\dagger} =\phi(f_{g}).$  
  \item[4] There exists a unique translation invariant vector, the vacuum vector $\Omega$, in $D.$
  \item[5] When the supports of $f$ and $g$ are spacelike separated, $\varphi(f)\varphi(g)$ coincides on $D$ with  $\varphi(g)\varphi(f).$
  \item[A1] The field operators $\varphi(f),$ for real-valued $f,$ are essentially self-adjoint \cite{Rudin:1991} on a dense common invariant domain $D.$
  \item[A2] When the supports of $f$ and $g$ are spacelike separated, $\left[\varphi(f),\varphi(g)\right]=0$,  meaning that the spectral projections of  $\varphi(f)$ commute with the spectral projections of  $\varphi(f)\varphi(g)$ \cite{Guido:2008jk}.
  
\end{description}
Properties \textbf{A1}  and \textbf{A2 } allow one to obtain a net of observable algebras under the axioms of Haag-Kastler, and along with the vacuum  \cite{Guido:2008jk}. \\\\The Bisognano-Wichmann theorem provides a connection between the geometric symmetries of  Minkowski spacetime and the algebraic structure of local observables. It associates the modular group of the algebra of observables with Lorentz boosts that preserve a wedge region. Also,   CPT reflections are linked to the modular conjugation (see \ref{TTT}).\\\\First let us  state the Bisognano-Wichmann result for a free scalar quantum field, the so called one-particle Bisognano-Wichmann theorem. The right Rindler wedge consists of ${\cal W}_R = \big\{ (x,y,z,t) \in {\mathbb{R}^4}, \; x> |t| \; \big\}.$ This region is invariant under the action of Lorentz boosts, given by the matrix,

$$\Lambda_{{\cal W}}(\tau) = \left[\begin{array}{cccc} \mathrm{cosh}(2\pi\tau) & -\mathrm{sinh}(2\pi\tau) & 0 & 0  \\ -\mathrm{sinh}(2\pi\tau) & \mathrm{cosh}(2\pi\tau) & 0 & 0 \\0 & 0 & 1 & 0 \\0 & 0 & 0 & 1 \end{array}\right]$$

\textbf{One-particle Bisognano-Wichman theorem}: \\

Let ${\cal H}_1 ({\cal W}_R)$ be the   one-particle space associated to the right Rindler wedge,  see Section \ref{fock1}.   ${\cal H}_1 ({\cal W}_R)$ is isomorphic to $L^2(H_m, d \mu_m)$, built out from test functions with support in ${\cal W}_R$.  It turns out that the modular operator and the modular conjugation have a geometric action on  the space of test functions $L^2(H_m, d \mu_m)$ given by 
\begin{equation} 
j_{{\cal W}}=\Theta \; u(R(\pi),0) \;, \qquad 
\delta^{i\tau}_{\cal W} = u(R(\Lambda_{{\cal W}}(\tau),0) \;. \label{lt}
\end{equation} 
In the above,  $u$ is a representation of the proper orthochronous Poincaré group acting on  elements $f \in L^2(H_m, d \mu_m)$ as $(u(\Lambda, a)f)(p) = {\rm e}^{i a . p} f(\Lambda^{-1} p,)$.  $\Theta$ is the CPT transformation and  $R$ denotes a rotation around the $x$ axis. \\\\{\underline {\bf Comment}} As it will be discussed in Sect \eqref{Sect6}, the modular operators $(j_{{\cal W}},\delta^{i\tau}_{\cal W})$ of eq.\eqref{lt} correspond to the lifting of the Tomita-Takesaki operators $(J_{\cal W},\Delta_{\cal W})$ to the space of test functions $L^2(H_m, d \mu_m)$. For instance, in the case of the Weyl operators $W_f = e^{i\varphi(f)}$, one has: 
\begin{equation} 
J_{\cal W} \;e^{i\varphi(f)} \; J_{\cal W} = e^{-i\varphi(j_{\cal W} f)} \;, \qquad \Delta_{\cal W} \;e^{i\varphi(f)} \; \Delta_{\cal W}^{-1} = e^{i\varphi(\delta_{\cal W} f)} \;.\label{lff1}
\end{equation}
In a more general frame, it is possible to restate the Reeh-Schlieder and Bisognano-Wichmann theorems making use of the G\aa rding- Wightman axioms (including {\bf A1} and {\bf A2}) .\\

\textbf{Reeh-Schlieder Theorem}:\\

For all non-empty open region $O$, the vacuum vector $\Omega$ is cyclic for the algebra ${\cal A}(O)$. When $O'$ is non- empty, $\Omega$ is a standard vector,  {\it i.e.} cyclic and separating, for ${\cal A}(O)$, see  \cite{Guido:2008jk}. \\

\textbf{Bisognano-WichmannTheorem}:\\

The action of the modular and  conjugation operators for the pair ${\cal A}({\cal W},\Omega)$ is, respectively,  $$J_{{\cal W}}=\Theta.U(R(\pi),0), $$ and
$$\Delta_{\cal W}^{i\tau} =  U(\Lambda_{{\cal W}}(\tau),0)$$ where  $U$ is a unitary representation of the orthochronous Poincaré group. \\\\ Moreover,   wedge duality ${\cal A}({\cal W}_R)'={\cal A}({\cal W_R'})$ holds. \\\\{\underline {\bf Comment}} One important consequence of the Bisognano-Wichmann Theorem is that the spectrum of the modular operator $\Delta_{\cal W}$ coincides with the positive real line ${\mathbb{R}}_{+}$, {\it i.e.} $\sigma_{sp}(\Delta_{\cal W}) = [0,\infty)$. This follows from the fact that the modular operator $\Delta_{\cal W}$ turns out to be related to the generator of the Lorentz boosts \cite{BW}. As such, the local algebras of a scalar field for wedge regions are of the type $III_1$, a feature which has deep consequences on the study of the entanglement in Quantum Field Theory \cite{Summers:1987ze}. The Bisognano-Wichmann results have been generalized to dilatation invariant theories for double cones regions, see \cite{Brunetti:1992zf} and refs. therein.   The modular operator $\Delta$ turns out to be related to the generator of the dilatations. Also here,  the spectrum of $\Delta$ is the positive real line $[0, \infty)$. Again, the local algebras are of type $III_1$. It is  a common  belief that the general structure of the local algebras of a Quantum Field Theory in Minkowski spacetime is of the type $III_1$ \cite{Buchholz:1986bg}.

\subsection{Dualities} 

The notion of "duality" between different regions of spacetime and their associated algebras of observables plays an important role in the scheme of AQFT.  Those duality principles help ensure the theory is in accordance to basic physical principles, such as locality, causality, and spacetime symmetries. 

\begin{description}
  \item[ Wedge duality]  This concept refers to the relationship between the algebra of observables localized in a Rindler wedge and the observables localized in the complementary wedge (the left Rindler wedge). In that case ${\cal A}({\cal W})'={\cal A}({\cal W'})$ is satisfied, and it makes sure that the algebra associated with the right Rindler wedge is the commutant of the algebra associated to its spacelike complement. 

  \item[ Haag duality ] This concept generalizes the notion of wedge duality, and it applies to more general spacetime regions, like double cones or bounded regions in Minkowski space. Here one has that the algebra of observables in a spacetime region should be equal to the commutant of the algebra associated with its causal complement. Given a region $O$ of  spacetime, we denote by $O'$ its causal complement, i.e., the region where all the points are spacelike separated from $O.$ Haag duality states that ${\cal A}(O)'={\cal A}({O'})$, such that the commutant of the algebra of observables localized in $O$ is equal to the algebra of observables localized in the causal complement $O'.$ It is interesting to see it as a special case of Einstein's causality \cite{Schroer2016} , which would be ${\cal A}(O)' \supseteq{\cal A}({O'}).$  Haag duality essentially states that all observables in the causal complement  of a region do not  just commute with those in the region itself but characterize the complement by means of the commutant structure.
  
  \item[ Essential duality ] Is a concept that deals with extending the local algebras in those cases where Haag duality fails, such that a weaker duality is recovered. Given a net $O \rightarrow  {\cal A}(O)$, $O$ being a double cone we define, for a general region $C$, ${\cal A}(C) = \lor_{O\subset C}{\cal A}(O).$ The dual net is defined as ${\cal A}(O)^{Dual} = {\cal A}(O')' .$ One says that the net satisfies essential duality if $O \rightarrow  {\cal A}(O)^{Dual}$  is local for double cones.

\end{description}

\section{ The Bell-CHSH inequality in Quantum Field Theory}\label{Sect6}	

\subsection{Generalities} 

We have now all ingredients for the formulation of the Bell-CHSH inequality in relativistic Quantum Field Theory. It is helpful to go back to Quantum Mechanics and briefly remind a few salient features of the Bell-CHSH inequality, see \cite{Guimaraes:2024byw} for a recent overview. \\\\One considers a bipartite system $AB$ of two spins $1/2$ where, as it is customary, the capital letters $A$ and $B$ refer to Alice and Bob. Taking as entangled state one of the four Bell states
\begin{equation} 
|\psi_{AB}\rangle = \frac{1}{\sqrt{2}} \left( |+ \rangle_A \otimes |+\rangle_B + |- \rangle_A \otimes |-\rangle_B\right) \;, \label{Bst}
\end{equation}
one looks at the correlation function 
\begin{equation} 
\langle \psi_{AB}|\; {\cal C}\;| \psi_{AB}\rangle = \langle \psi_{AB}|\; (A+A') \otimes B + (A-A') \otimes B'\;| \psi_{AB}\rangle \;, \label{qmc}
\end{equation} 
where $(A,A',B,B')$ are Hermitian dichotomic operators fulfilling the conditions 
\begin{eqnarray} 
A & =  & A^{\dagger} \;, \qquad A^2 =1 \;, \qquad A'= A'^{\dagger} \;, \qquad A'^2 =1 \;, \nonumber \\
B& =  & B^{\dagger} \;, \qquad B^2 =1 \;, \qquad B'= B'^{\dagger} \;, \qquad B'^2 =1 \;, \label{1c}
\end{eqnarray}
and 
\begin{equation} 
[A, B]  =   0 \;, \qquad  [A, B'] =0 \;, \qquad [A',B]=0 \;, \qquad [A',B']=0 \;, \label{2c} 
\end{equation}
\begin{equation}
[A, A]  \neq   0 \;, \qquad [B,B'] \neq 0 \;. \label{3c}
\end{equation} 
Setting 
\begin{eqnarray} 
A |+\rangle_A & =&  e^{i\alpha} |-\rangle_A \;, \qquad A |-\rangle_A = e^{-i\alpha} |+\rangle_A \;, \qquad A'|+\rangle_A = e^{i\alpha'} |-\rangle_A\;, \qquad A' |-\rangle_A = e^{-i\alpha'} |+\rangle_A \;, \nonumber \\
B |+\rangle_B & =&  e^{i\beta} |-\rangle_B \;, \qquad B |-\rangle_B = e^{-i\beta} |+\rangle_B \;, \qquad B'|+\rangle_B = e^{i\beta'} |-\rangle_B\;, \qquad B' |-\rangle_B = e^{-i\beta'} |+\rangle_B \;, \label{aabb}
\end{eqnarray} 
where $(\alpha,\alpha',\beta,\beta')$ are arbitrary parameters, one gets 
\begin{equation} 
\langle \psi_{AB}|\; {\cal C}\;| \psi_{AB}\rangle  = \cos(\alpha+\beta) + \cos(\alpha'+\beta)+ \cos(\alpha+\beta')- \cos(\alpha'+\beta') \;. \label{rc}
\end{equation}
One speaks of a violation of the Bell-CHSH inequality whenever 
\begin{equation} 
2 < \big| \langle \psi_{AB}|\; {\cal C}\;| \psi_{AB}\rangle \big| \le 2 \sqrt{2} \;. \label{vc}
\end{equation}
The value $ 2 \sqrt{2}$, known as Tsirelson's bound \cite{tsi1}, is the maximum value attainable by the correlator  \eqref{Bst}. Choosing 
\begin{equation}
\alpha = 0 \;, \qquad \alpha'=\frac{\pi}{2} \;, \qquad \beta=-\frac{\pi}{4}\;, \qquad \beta'= \frac{\pi}{4} \;, \label{vang}
\end{equation} 
one gets 
\begin{equation} 
\langle \psi_{AB}|\; {\cal C}\;| \psi_{AB}\rangle  = 2 \sqrt{2} \;, \label{mv}
\end{equation}
showing that the Bell state $|\psi_{AB}\rangle$ gives maximal violation. 

\subsection{The Summers-Werner Theorem for wedge regions for a free scalar field} 

In order to formulate the Bell-CHSH inequality in Quantum Field Theory, one starts by considering two spacelike separated regions ${\cal O}_1$ and ${\cal O}_2$ in Minkowski spacetime and a set of bounded Hermitian field operators  \cite{Summers:1987fn,Summ}
\begin{eqnarray} 
A &=& A^{\dagger} \;, \qquad ||A|| \le 1 \;, \qquad A'= A'^{\dagger} \;, \qquad ||A'|| \le 1 \;, \nonumber \\
B &=& B^{\dagger} \;, \qquad ||B|| \le 1 \;, \qquad B'= B'^{\dagger} \;, \qquad ||B'|| \le 1 \;, \label{qftop}
\end{eqnarray} 
where $|| \cdot ||$ stands for the norm operator and $(A,A') \in {\cal A}({\cal O}_1)$, $(B,B') \in {\cal A}({\cal O}_2)$, with ${\cal A}({\cal O}_1)$,  ${\cal A}({\cal O}_2)$ von Neumann algebras obtained from a net of local algebras $\{ {\cal A}({\cal O}) \}$
\begin{equation} 
{\cal O} \rightarrow {\cal A}({\cal O}) \;, \label{ntt}
\end{equation}
obeying the Haag-Kastler axioms, see Sect.\eqref{HK}. \\\\An example of ${\cal A}({\cal O})$ is provided by the von Neumann algebra built out from Weyl operators of a real scalar field $\varphi(h)$, namely 
\begin{equation} 
{\cal A}({\cal O}) = \big\{ e^{i \varphi(h)}, \; h \in {\cal S}({\mathbb R}^4), \; supp(h) 	\subseteq {\cal O} \big\}'' \;, \label{bbcc}
\end{equation}
where ${'}{'}$ means the bicommutant. \\\\Let $\phi$ be a state on the Von Neumann algebra ${\cal A}$. One introduces the Bell-CHSH correlation function\footnote{Notice that a slightly different normalization has been adopted for $\beta \big[\phi, {\cal A}({\cal O}_1) ,{\cal A}({\cal O}_2) \big]$ with respect to the original work   \cite{Summers:1987fn,Summ}.}    \cite{Summers:1987fn,Summ}
\begin{equation} 
\beta \big[\phi, {\cal A}({\cal O}_1) ,{\cal A}({\cal O}_2) \big] \equiv \; sup \left[ \phi \left( (A+A')B+(A-A')B' \right) \right] \;, \label{betasw}
\end{equation}
where the supremum is taken over all $(A,A') \in  {\cal A}({\cal O}_1)$ and $(B,B') \in  {\cal A}({\cal O}_2)$. \\\\It follows  \cite{Summers:1987fn,Summ} that if ${\cal O}_1 \subseteq {\cal O}_2'$  where ${\cal O}_2'$ is the causal complement of ${\cal O}_2$, then 
\begin{equation} 
\beta \big[\phi, {\cal A}({\cal O}_1) ,{\cal A}({\cal O}_2) \big] \le 2 \sqrt{2} \;, \label{qftts}
\end{equation}
expressing the Tsirelson bound in Quantum Field Theory. \\\\We remind that, given a region ${\cal O}$ in Minkwsky spacetime, its causal complement ${\cal O}'$ is 
\begin{equation}
{\cal O}'= \{ y \in {\mathbb{R}^4},\; (y-x)^2<0,\; x\in {\cal O} \}  \;. \label{cauc}
\end{equation}
Similarly to Quantum Mechanics, one speaks of a violation of the Bell-CHSH inequality whenever  \cite{Summers:1987fn,Summ}
\begin{equation} 
2 <  \beta \big[\phi, {\cal A}({\cal O}_1) ,{\cal A}({\cal O}_2) \big]   \le 2 \sqrt{2} \;. \label{vqft}
\end{equation}
Before going any further, a few remarks are in order. As it stands, the analysis of the correlation function \eqref{betasw} requires the specification of several issues:
\begin{itemize} 
\item two spacelike separated regions ${\cal O}_1$ and ${\cal O}_2$ have to be specified. This ensures that Alice's measurements do not affect Bob's measurements and vice versa. In what follows, the regions ${\cal O}_1$ and ${\cal O}_2$ will be identified with the right and left Rindler wedges, ${\cal O}_1  \leftrightarrow {\cal W}_R$ and ${\cal O}_2 \leftrightarrow {\cal W}_L$
\begin{equation} 
{\cal W}_R = \big\{ (x,y,z,t) \in {\mathbb{R}^4}, \; x > |t| \; \big\}\;, \qquad {\cal W}_L = \big\{ (x,y,z,t) \in {\mathbb{R}^4}, \; -x > |t| \; \big\}\;.  \label{rdw}
\end{equation}
These regions are left invariant by the Lorentz boosts. Moreover, they are the causal complement of each other. 
\item concerning the field content, the Bell-CHSH correlation function \eqref{betasw} will be referred to a scalar real massive field in $1+3$ Minkowski spacetime. 
\item It remains to specify the state $\phi$, which will be taken to be to vacuum state $|0 \rangle$
\begin{equation} 
\phi \rightarrow \phi_0\;, \qquad \phi_0(A) = \langle 0| \; A |\; 0 \rangle \;, A \in {\cal A} \;. \label{phi0} 
\end{equation}
\end{itemize}
With these specifications, we can state the result obtained by Summers-Werner \cite{Summers:1987fn,Summ}\\\\{\bf {\underline{Theorem}}}
\begin{equation} 
\beta \big[\phi_0, {\cal A}({\cal W}) ,{\cal A}({\cal W}') \big] \equiv \; sup  \left[  \phi_0 \left( (A+A')B+(A-A')B' \right) \right]= 2 \sqrt{2}  \;, \label{swt}
\end{equation}
where $(A,A') \in {\cal A}(\cal W)$, $(B,B') \in {\cal A}({\cal W}')$ and ${\cal W}$ is any wedge region \footnote{ ${\cal W}$ denotes any region obtained by means of a Poincar{\'e} transformation of ${\cal W}_R$, eq.\eqref{rdw}. }.\\\\Notice that, from Haag's duality, one has 
\begin{equation} 
{\cal A}({\cal W}') = {\cal A}'({\cal W}) \;. \label{hdlWW}
\end{equation} 
The region ${\cal W}'$ stands for the causal complement of ${\cal W}$. One says that $({\cal W},{\cal W}')$ are complementary wedges. 
This remarkable theorem means that, given two complementary wedges $({\cal W},{\cal W'})$, there exists a quadruple of bounded Hermitian operators $(A,A',B,B')$ giving maximal violation of the Bell-CHSH inequality in the vacuum state $|0\rangle$. It is worth underlining that this result holds at the level of free fields, a feature which strengthen the unavoidable manifestation of the entanglement in relativistic Quantum Field Theory. \\\\Although stated for a real scalar field, the above Theorem has been proven also f/Users/silviopaolosorella/Desktop/Screen Shot 2025-04-08 at 06.43.14.pngor free spinor fields   
\cite{Summers:1987fn,Summ}. A generalization to other states than the vacuum state can be found in \cite{Summers:1987ze}. 

\subsection{Details of the proof of the Summers-Werner Theorem}\label{detSW}

Let us reproduce here the main steps of the aforementioned Theorem, as given in \cite{Summers:1987fn,Summ} . One first shows that 
\begin{equation} 
\beta \big[\phi_0, {\cal A}({\cal W}) ,{\cal A}({\cal W}') \big] \ge 2\sqrt{2} \;\frac{2\lambda}{1+\lambda^2} \;, \label{dlb}
\end{equation}
where $\lambda \in [0,1]$. \\\\Then, ones proceeds by proving that  $\lambda=1$ is a non-trivial accumulation point of the spectrum of the Tomita-Takesaki modular operator $\Delta_{\cal W}$. \\\\The above points require the use of all ingredients presented in the previous sections, namely: von Neumann algebras, Tomita-Takesaki modular theory, Bisognano-Wichmann results. \\\\Concerning the first point one picks up two set of test functions $(f,f')$ and $(g,g')$, supported, respectively, in ${\cal W}$ and ${\cal W'}$. Let now ${\tilde M}(f,f')$ and $ {\tilde N}(g,g')$ denote the subspaces generated by the linear span of $(f,f')$ and $(g,g')$. It is apparent that 
\begin{equation} 
\beta \big[\phi_0, {\cal A}({\cal W}) ,{\cal A}({\cal W}') \big] \ge \beta \big[\phi_0, {\cal A}({\tilde M}(f,f')) ,{\cal A}( {\tilde N}(g,g')) \big] \;, \label{lb1}
\end{equation}  
which follows by observing that ${\tilde M}(f,f')$ and $ {\tilde N}(g,g')$ are finite dimensional test functions subspaces. Moreover, as long as one works with finite dimensional subspaces, the renowed von Neumann Theorem on the unitary equivalence of the representations of the canonical commutation relations applies, meaning that ${\cal A}({\tilde M}(f,f'))$ and ${\cal A}( {\tilde N}(g,g'))$ are unitary equivalent to ${\cal B}({\cal H}_1)$ and to ${\cal B}({\cal H}_2)$
\begin{equation} 
{\cal A}({\tilde M}(f,f')) \cong {\cal B}({\cal H}_1) \;, \qquad {\cal A}( {\tilde N}(g,g')) \cong {\cal B}({\cal H}_2) \;, \label{lb2}
\end{equation}
where ${\cal H}_1$ and ${\cal H}_2$ are Hilbert spaces carrying a representation of the canonical commutation relations for one degree of freedom, namely 
\begin{eqnarray}
{\cal H}_1 & \Leftrightarrow&  (a,a^\dagger) \;, \qquad [a,a^\dagger]=1 \;, \nonumber \\
{\cal H}_2 & \Leftrightarrow&  (b,b^\dagger) \;, \qquad \;\;[b,b^\dagger]=1 \;. \label{lb3}
\end{eqnarray} 
At this stage, one proceeds by showing that, on $ {\cal H}_1 \otimes {\cal H}_2$, it is always possible to construct a Bell-CHSH inequality such that
\begin{equation} 
\langle \Omega |\; {\cal C} \; | \Omega \rangle_{{{\cal H}_1} \otimes{ {\cal H}_2}} = \langle \Omega |\; (A+A')\otimes B + (A-A')\otimes B'|;\Omega\rangle = 2\sqrt{2} \;\frac{2\lambda}{1+\lambda^2} \;, \label{lb4}
\end{equation} 
where $(A,A')$ and $(B,B')$ are Hermitian dichotomic operators acting on ${\cal H}_1$ and ${\cal H}_2$, while $|\Omega\rangle$ stands for the fundamental state of some suitable Hamiltonian on $ {\cal H}_1 \otimes {\cal H}_2$. \\\\As outlined in \cite{Summers:1987fn,Summ}, the state $\Omega$ can be identified with the two-mode squeezed states of two Harmonic oscillators, namely 
\begin{equation} 
|\Omega \rangle = (1-\lambda^2)^{1/2} \sum_{n=0}^\infty \lambda^n |n_1 n_2\rangle \;, \qquad |n_1 n_2\rangle = \frac{1}{n!} (a^\dagger)^n (b^\dagger )^n |0\rangle_F \;, \label{lb5}
\end{equation}
with $|0\rangle_F$ being the Fock vacuum of the two oscillators system 
\begin{equation} 
H_0 = (a^\dagger a + b^\dagger b) \;, \label{lb6}
\end{equation}
It turns out that the squeezed state $!\Omega\rangle$ can be cast in the form 
\begin{equation} 
!\Omega\rangle = {\cal S}(\lambda)  |0\rangle_F \;, \label{lb6}
\end{equation} 
where ${\cal S}(\lambda) $ is a unitary operator, {\it i.e.} ${\cal S}(\lambda){\cal S}^\dagger(\lambda) ={\cal S}^\dagger(\lambda) {\cal S}(\lambda)=1 $, namely:
\begin{equation} 
{\cal S}(\lambda) = e^{\tanh(r) a^\dagger b^\dagger} \; e^{-\log(\cosh(r)) (a^\dagger a +b^\dagger b + 1)} \; e^{-\tanh(r) ab}  \; \qquad \lambda = \tanh(r) \;, \qquad r\ge 0. 
\end{equation} 
As a consequence, the state $|\Omega\rangle$ is the fundamental state of the Bogoliubov type Hamiltonian  \cite{Summers:1987fn,Summ} ${\tilde H}$ given by 
\begin{equation} 
{\tilde H} = {\cal S}\;(\lambda) H_{0}\;{\cal S}^\dagger(\lambda) \;, \qquad {\tilde H} |\Omega\rangle =0 \;. \label{lb7}
\end{equation}
Concerning now the Bell-CHSH inequality \eqref{lb4}, for the operators $(A,A',B,B')$ one has  \cite{Summers:1987fn,Summ}
\begin{eqnarray} 
A |2n\rangle_1 & = & e^{i \alpha} |2n+1\rangle_1 \;, \qquad A |2n+1\rangle_1  =  e^{-i \alpha} |2n\rangle_1 \;, \nonumber \\
A' |2n\rangle_1 & = & e^{i \alpha'} |2n+1\rangle_1 \;, \qquad A' |2n+1\rangle_1  =  e^{-i \alpha'} |2n\rangle_1 \;, \nonumber \\
B |2n\rangle_2 & = & e^{i \beta} |2n+1\rangle_2 \;, \qquad B |2n+1\rangle_2  =  e^{-i \beta} |2n\rangle_2 \;, \nonumber \\
B' |2n\rangle_2 & = & e^{i \beta'} |2n+1\rangle_2 \;, \qquad B' |2n+1\rangle_2  =  e^{-i \beta'} |2n\rangle_2 \;, \label{lb8}
\end{eqnarray}
with $(\alpha, \alpha', \beta, \beta')$ arbitrary parameters. After a simple algebra, one finds 
\begin{equation} 
\langle \Omega|\; A\otimes B\;|\Omega\rangle = \frac{2 \lambda}{1+\lambda^2} \cos(\alpha+\beta) \;, \label{lb9}
\end{equation}
so that 
\begin{equation} 
\langle \Omega |\; {\cal C} \; | \Omega \rangle_{{{\cal H}_1} \otimes{ {\cal H}_2}} = \frac{2 \lambda}{1+\lambda^2}\left(  \cos(\alpha+\beta) + \cos(\alpha'+\beta)+ \cos(\alpha+\beta')- \cos(\alpha'+\beta') \right) \;. \label{lb10}
\end{equation}
Finally, setting 
\begin{equation} 
\alpha = 0 \;, \qquad \alpha'=\frac{\pi}{2} \;, \qquad \beta=-\frac{\pi}{4}\;, \qquad \beta'=\frac{\pi}{4} \;, \label{lb11}
\end{equation} 
one finds the desired result:
\begin{equation} 
\langle \Omega |\; {\cal C} \; | \Omega \rangle_{{{\cal H}_1} \otimes{ {\cal H}_2}} = 2 \sqrt{2} \frac{2 \lambda}{1+\lambda^2}  \;. \label{lb12}
\end{equation}
To complete the proof, it remains to show that the point $\lambda=1$ is in the spectrum of the Tomita-Takesaki modular operator $\Delta_{\cal W}$. This is assured by the Bisognano-Wichmann results, stating that the spectrum of $\Delta_{\cal W}$ coincides with the positive real line ${\mathbb{R}}_{+}$, see subsect. \eqref{bwth}. The value $\lambda=1$ can thus be achieved, completing the proof of the Summers-Werner Theorem for complementary wedges regions. 

\subsection{The correlation function of Weyl operators}

We shall now attempt  at  providing a check of the Summers-Werner theorem stated previously. The task is far from being trivial due to several reasons. Perhaps, the most difficult one is that, in the case of a scalar field, the explicit expression of the bounded Hermitian operators $(A,A',B,B')$  leading to the maximal violation of the Bell-CHSH inequality is actually not at our disposal. Though, it is not difficult to figure out examples of bounded Hermitian operators. suitable for the study of the Bell-CHSH inequality.  For instance, due to the properties of the hyperbolic tangent, the operator 
\begin{equation}  
{\cal Q}_h = \tanh(\varphi(h))  \;, \label{th}
\end{equation}
would provide an example of a Hermitian bounded operator. Other examples are provide by 
\begin{equation}  
{\hat {\cal Q}}_h = e^{-\varphi(h)^2}  \;, \label{th1}
\end{equation}
and 
\begin{equation}  
{\tilde {\cal Q}}_h = \frac{1}{1+ \varphi(h)^2}  \;, \label{th2}
\end{equation}
As we shall illustrate in the following, the correlation functions of these bounded operators can be evaluated in a very powerful and elegant way by means of the Weyl operators, see subsection\eqref{dich}. \\\\It is thus helpful focusing first on the unitary Weyl operators $W_h$. Accordingly, we shall consider the Bell-CHSH type correlation  function 
\begin{equation} 
\langle 0|\; {\cal C}\; |0\rangle = \langle 0|\; (A_f+A_{f'})B_g + (A_f-A_{f'})B_{g'}\; |0\rangle \;, \label{Wcc}
\end{equation}
with
\begin{equation} 
A_f = W_f = e^{i \varphi(f)}\;, \qquad A_{f'} = W_{f'} = e^{i \varphi(f')}\; \qquad B_g = W_g = e^{i \varphi(g)}\;\qquad B_{g'} = W_{g'} = e^{i \varphi(g')}\;, \label{opw}
\end{equation}  
and 
\begin{equation} 
supp(f,f') \subseteq {\cal W}_R \;, \qquad supp(g,g') \subseteq {\cal W}_L  \;, \label{ffgg}
\end{equation}
so that, as required by eqs.\eqref{2c},\eqref{3c}: 
\begin{equation} 
[A_f,B_g]=[A_f, B_{g'}] = [A_{f'}, B_g] = [A_{f'},B_{g'}] = 0 \;. \label{req}
\end{equation}
Several justifications can be given for the use of the correlation function \eqref{Wcc}. For instance, as discussed in \cite{Guimaraes:2024alk}, one can rely on the generalization of the Bell-CHSH inequality to unitary operators. A second way \cite{Guimaraes:2024lqf} of deriving expression \eqref{Wcc} is that of working on the enlarged space ${\cal H} \otimes {\cal H}_{AB}$, where ${\cal H}$ is the Hilbert space of the scalar quantum field and ${\cal H}_{AB}$ is the Hilbert space of a bipartite quantum mechanical system made up by two spins $1/2$: ${\cal H}_{AB}= {\cal H}_A \otimes {\cal H}_B$. Working on ${\cal H} \otimes {\cal H}_{AB}$, it turns out to be easy to introduce Hermitian dichotomic operators and construct a Bell-CHSH inequality in ${\cal H} \otimes {\cal H}_{AB}$. Afterwords, the result is projected onto ${\cal H}$, thus obtaining precisely the correlation function  \eqref{Wcc}. Due to its simplicity, it is worth giving a summary of this procedure.

\subsubsection{Working in ${\cal H} \otimes {\cal H}_{AB}$: construction of Hermitian  dichotomic operators} 

Following \cite{Guimaraes:2024lqf}, we shall consider  a system whose Hilbert space is ${\cal H} \otimes {\cal H}_{AB}$, where ${\cal H}$ is the Hilbert space of the scalar quantum field and ${\cal H}_{AB}$ is the Hilbert space of a bipartite quantum mechanical system made up by two spins $1/2$: ${\cal H}_{AB}= {\cal H}_A \otimes {\cal H}_B$.  \\\\The first task is that of constructing a von Neumann algebra on ${\cal H} \otimes {\cal H}_{AB}$ with a cyclic and separating state. As we have seen,  the Hilbert space ${\cal H}$ reads
\begin{equation} 
{\cal H} = {\rm span}  \{ W_{f_1}.....W_{f_n} |0\rangle \} \label{h}
\end{equation}
where $W_f = e^{i\varphi(f)}$ stands for the Weyl operator and $(f_1,...,f_n)$ are test functions.  The vacuum state $|0\rangle$ is a cyclic and separating state for the Von Neumann Algebra generated by the Weyl operators $\{ W_f \}$. As for the cyclic and separating state on ${\cal H}_{AB}$, the goal is accomplished by considering the maximally entangled Bell state $|\psi_{AB}\rangle$ of eq. \eqref{Bst}. This state  is cyclic and separating for the von Neumann algebra 
\begin{equation} 
{\hat {\cal O}} =\{ O_A \otimes {\mathbb{1}}_B \}  \;, \label{vno}
\end{equation} 
where $\{ O_A \}$ are the operators of ${\cal H}_A$\footnote{The operators $\{O_A \}$ are the $2 \times 2 $ complex matrices.}. \\\\Therefore, the state 
\begin{equation} 
|\psi \rangle = |0 \rangle \otimes |\psi_{AB} \rangle \;, \label{psi} 
\end{equation} 
is cyclic and separating for the von Neumann algebra 
\begin{equation} 
{\hat {\cal A}_O} = {\cal A} \otimes {\hat {\cal O}}  \;. \label{vnao}
\end{equation}
One can now introduce the operators $(A,A'.B.B')$ defined by 
\begin{equation}
A | + \rangle_A = e^{i\varphi(f)}\;|-\rangle_A  \;, \qquad  A | - \rangle_A = e^{-i\varphi(f)}\; | + \rangle_A  \;, \label{aop}
\end{equation}
\begin{equation} 
A' | + \rangle_A = e^{i\varphi(f')}\;|-\rangle_A  \;, \qquad  A' | - \rangle_A = e^{-i\varphi(f')}\; |+\rangle_A  \;, \label{apop}
\end{equation}
where $(f,f')$ denote Alices's test functions. Similarly
\begin{equation} 
B | + \rangle_B = e^{i\varphi(g)}\;|-\rangle_B  \;, \qquad  B | - \rangle_B = e^{-i\varphi(g)}\; |+\rangle_B  \;, \label{bop}
\end{equation}
and
\begin{equation} 
B' | + \rangle_B = e^{i\varphi(g')}\;|-\rangle_B  \;, \qquad  B' | - \rangle_B = e^{-i\varphi(g')}\; |+\rangle_B  \;, \label{bpop}
\end{equation}
where $(g,g')$ are Bob's test functions. \\\\One easily checks that the operators $(A,A')$, $(B,B')$ fulfill all needed requirements, that is:
\begin{eqnarray} 
A^{\dagger} & = & A \;, \qquad A^2 =1 \;, \qquad A'^{\dagger} = A' \;, \qquad A'^2=1 \nonumber \\
B^{\dagger} & = & B \;, \qquad B^2 =1 \;, \qquad B'^{\dagger} = B' \;, \qquad B'^2=1 \;, \label{abb}
\end{eqnarray} 
with 
\begin{equation}
 [  A ,B]  = [A,B']= [A',B]= [A,B'] = 0 \;. \label{abb1}
\end{equation} 
The quadruple $(A,A',B,B')$ is thus an eligible set of operators for the Bell-CHSH inequality in ${\cal H} \otimes {\cal H}_{AB}$, namely 
\begin{equation} 
\langle {\cal C}_{{\cal H} \otimes {\cal H}_{AB}} \rangle_{\psi} = \langle \psi | \; (A + A') \otimes B + (A-A') \otimes B' \; | \psi \rangle \;. \label{cc1}
\end{equation}
Since $(A,A',B,B')$ are Hermitian and dichotomic, Tsirelson's argument \cite{tsi1} applies, implying that 
\begin{equation} 
|| {\cal C}_{{\cal H} \otimes {\cal H}_{AB}} || \le 2 \sqrt{2} \;, \label{ts}
\end{equation} 
Let us now consider the operator ${\cal C}_{\cal H}$ acting on the Hilbert space of the quantum field ${\cal H}$, obtained by working out the expectation value of ${\cal C}_{{\cal H} \otimes {\cal H}_{AB}}$ in the Hilbert space  ${\cal H}_{AB}$ of the quantum mechanical system, {\it i.e.} 
\begin{equation} 
{\cal C}_{\cal H} = \langle \psi_{AB} \; | {\cal C}_{{\cal H} \otimes {\cal H}_{AB}} \;| \psi_{AB} \rangle \;. \label{cnot}
\end{equation}
From equation \eqref{ts}, it turns out that 
\begin{equation} 
| \langle 0 | \; {\cal C}_{\cal H} \; |  0 \rangle  | \le 2 \sqrt{2} \;. \label{tss}
\end{equation}
Moreover, an elementary calculation shows that 
\begin{equation} 
\langle \psi_{AB} \;| A\otimes B \;| \psi_{AB} \rangle = \cos(\varphi(f) + \varphi(g)) \;. \label{cs1}
\end{equation} 
Therefore, taking into account that the vacuum expectation value of an odd number of fields vanishes, for the Bell-CHSH of the quantum field in the vacuum state, we get 
\begin{equation} 
\langle 0| \; {\cal C}_{\cal H} \;| 0 \rangle = \langle 0|\; (W_f+ W_{f'})W_{g} + (W_f -W_{f'})W_{g'} \;|0\rangle =\langle 0|\; ( e^{i\varphi(f)} + e^{i\varphi(f')} ) 
e^{i \varphi(g) }  +  (e^{i\varphi(f)} - e^{i\varphi(f')} )  e^{i \varphi(g) }\; |0\rangle  \label{bco}
\end{equation}
A violation of the Bell-CHSH inequality in the vacuum state occurs if  
\begin{equation} 
2 < | \langle 0|\; {\cal C}_{\cal H}\;|0 \rangle | \le 2 \sqrt{2} \;. \label{vb}
\end{equation} 
Expression \eqref{bco}  is nothing but he correlation function of eq.\eqref{Wcc}.

\subsubsection{Lifting the Tomita-Takesaki modular theory to the space of test functions} 

Before proceeding with the analysis of the violation exhibited by the correlation function \eqref{Wcc}, it is worth presenting an important aspect of the modular theory of Tomita-Takesaki related to the Weyl operators. For these operators, the modular structure can be lifted to the space of test functions, allowing the evaluation of the correlator \eqref{Wcc} in closed form. Although we are referring to the complementary wedges $({\cal W}_R, {\cal W}_L)$, eq.\eqref{rdw}, the presentation can be given at more general level. A detailed account of this topic can be found in  \cite{Summers:1987fn,Summ,Guido:2008jk}. \\\\The possibility of lifting the action of the modular operators $(\Delta,J)$ stems from the following result \cite{RD}. Let ${\cal A}$ be a von Neumann algebra on a Hilbert space ${\cal H}_{\cal A}$, equipped with a cyclic and separating vector $|\omega\rangle$. Let ${\cal A}_s$ be the set of the self-adjoint elements of ${\cal A}$: $A\in {\cal A}_s \Rightarrow A^{\dagger}=A$. It follows that the subspace 
\begin{equation} 
{\cal K} = {\rm span} \{ {\cal A}_s |\omega\rangle \} \;, \label{subk}
\end{equation} 
is a standard real subspace for ${\cal H}_{\cal A}$, meaning that 
\begin{eqnarray} 
{\cal K} \cap  i {\cal K} &=& \{ 0 \} \;, \nonumber \\
{\cal K} + i {\cal K}& &{\rm is\; dense \; in\;} {\cal H}_{\cal A} \;. \label{rdr}
\end{eqnarray}
Let us observe that, on self-adjoint elements, $A=A^{\dagger}$, the action of the operator $S$ reduces to 
\begin{equation} 
S A |\omega\rangle = A |\omega\rangle \;. \label{obs}
\end{equation}
The lifting of $(S, \Delta,J)$ to the space of test functions follows thus by reminding that, as discussed in Sect.\eqref{Hilb}, 
the 1-particle space ${\cal H}_1$ is isomorphic to the Hilbert space $L^2(H_m, d \mu_m)$ of test functions \cite{Guido:2008jk}. \\\\More precisely, following  \cite{Summers:1987fn,Summ}, let $\mathcal{O}$ be an open region of the Minkowski spacetime and let $\mathcal{M}(\mathcal{O})$ be the real vector space of smooth test functions $\in \mathcal{C}_{0}^{\infty}(\mathbb{R}^4)$ with support contained in $\mathcal{O}$:
\begin{align} 
	\mathcal{M}(\mathcal{O}) = \{ f \, \vert supp(f) \subseteq \mathcal{O} \}. \label{MO}
\end{align}
One introduces the symplectic complement \cite{Summers:1987fn,Summ}  of $\mathcal{M}(\mathcal{O})$ as 
\begin{align} 
	\mathcal{M'}(\mathcal{O}) = \{ g \, \vert  \Delta_{PJ}(g,f) = 0, \; \forall f \in \mathcal{M}(\mathcal{O}) \}, \label{MpO}
\end{align}
that is, $\mathcal{M}(\mathcal{O})$ is given by the set of all test functions for which the smeared Pauli-Jordan expression $\Delta_{PJ}(f,g)$  
\begin{equation} 
\left[ \varphi(f), \varphi(g) \right] =  i  \Delta_{PJ}(f,g) \;. \label{smpj}
\end{equation} 
vanishes for any $f$ belonging to $\mathcal{M}(\mathcal{O})$. The symplectic complement $\mathcal{M'}(\mathcal{O})$ allows one to recast causality as \cite{Summers:1987fn,Summ}
\begin{align}
    \left[ \varphi(f), \varphi(g) \right] = 0,
\end{align}
whenever $f \in \mathcal {M}(\mathcal{O})$ and $g \in \mathcal {M'}(\mathcal{O})$. \\\\Let us denote by $\mathcal{A}(\mathcal{M})$
the von Neumann algebra made up by the Weyl operators $\{ W_f = e^{i \varphi(f)}\;, f\in {\cal M}\}$.  From the Reeh-Schlieder theorem \cite{Reeh:1961ujh,Haag:1992hx}, it turns out that the vacuum state $\vert 0 \rangle$ is cyclic and separating for  $\mathcal{A}(\mathcal{M})$, so that one can make use of the Tomita-Takesaki modular theory. \\\\As already underlined, this theory is particularly suited for analyzing the Bell-CHSH inequality within relativistic Quantum Field Theory  \cite{Summers:1987fn,Summ}. As discussed in  \cite{DeFabritiis:2023tkh},  it gives a way of constructing in a purely algebraic way Bob's operators from Alice's ones by using the modular conjugation $J$. That is, given Alice's operator $A_f$, one can assign the operator $B_f = J A_f J$ to Bob, with the guarantee that they commute since, by the Tomita-Takesaki theorem, the operator $B_f = J A_f J$ belongs to the commutant $\mathcal{A'}(\mathcal{M})$ \cite{DeFabritiis:2023tkh}. \\\\When equipped with the Lorentz-invariant inner product $\langle f \vert g\rangle$, eq.\eqref{scpd}, the set of test functions gives rise to a complex Hilbert space ${\cal H}_{\cal M}$ which enjoys several features. It turns out that the subspace $\mathcal{M}$ is a real standard   subspace for ${\cal H}_{\cal M}$ \cite{RD}:  $\mathcal{M} \cap i \mathcal{M} = \{ 0 \}$;  $\mathcal{M} + i \mathcal{M}$ is dense in ${\cal H}_{\cal M}$. From \cite{RD}, for such subspaces it is possible to set a modular theory analogous to that of the Tomita-Takesaki. One introduces an operator $s$ acting on $\mathcal{M} + i\mathcal{M}$ as
\begin{align}
    s (f+ih) = f-ih. \;, 
    \label{saction}
\end{align}
for $f,h \in \mathcal{M}$. Notice that $s^2 = 1$. Using the  polar decomposition, one has:  
\begin{align}
    s = j \delta^{1/2},  \label{jd}
\end{align}
where $j$ is an anti-unitary operator and $\delta$ is  positive and self-adjoint.  Similarly to the operators $(J,\, \Delta)$, the  operators $(j,\, \delta)$ fulfill  the following properties \cite{RD}:
\begin{align}
    j \delta^{1/2} j &= \delta^{-1/2}, \,\,\,\,\,\,  \delta^\dagger = \delta\nonumber \\
    s^\dagger &= j \delta^{-1/2} \;. \label{sdagger}
    \end{align}
Moreover,  a test function $f$ belongs to $\mathcal{M}$ if and only if \cite{RD}
\begin{equation} 
s f = f \;. 
    \label{sff}
\end{equation}
One observes that, due to the isomorphism existing between the 1-particle space ${\cal H}_1$ and the space of test functions, the content of eq.\eqref{sff} is the same as that of eq.\eqref{obs}. \\\\Similarly,  one has that $f' \in \mathcal{M}'$ if and only if $s^{\dagger} f'= f'$. \\\\The lifting of the action of the operators $(J, \Delta)$ to the space of test functions is implemented by  \cite{eck} 
\begin{align} 
 J e^{i {\varphi}(f) } J  = e^{-i {\varphi}(jf) }, \quad \Delta e^{i {\varphi}(f) } \Delta^{-1} = e^{i {\varphi}(\delta f) }. \label{jop}
\end{align} 
Also, it is worth noting that if $f \in \mathcal{M} \implies jf \in \mathcal{M}'$. This property follows from 
\begin{equation} 
s^{\dagger} (jf) = j \delta^{-1/2} jf = \delta^{1/2} f = j (j\delta^{1/2} f) = j (sf) = jf \;. \label{jjf} 
\end{equation} 
Making use of the operators $(j,\delta)$, for Alice's and Bob's Weyl operators we write 
\begin{eqnarray} 
A_f & =& e^{i \varphi(f)} \;, \qquad A_{f'} = e^{i\varphi(f')} \;, \qquad (f,f') \in {\cal M} \;, \nonumber \\
B_{jf} & =& e^{i \varphi(jf)} \;, \qquad b_{jf'} = e^{i\varphi(jf')} \;, \qquad (jf,jf') \in {\cal M'} \;. \label{abjo}
\end{eqnarray}
As required, the operators $(B_{jf},B_{jf'})$ commute with $(A_f,A_{f'})$ as $(B_{jf},B_{jf'})$ belong to the commutant of $\mathcal{A}(\mathcal{M})$. Therefore, for the correlation function \eqref{Wcc}, one gets 
\begin{eqnarray} 
\langle 0|\; {\cal C} \;|0\rangle & =&  \langle 0|\; e^{i\varphi(f+jf)} + e^{i\varphi(f'+jf)}+ e^{i\varphi(f+jf')} - e^{i\varphi(f'+jf')}\;|0\rangle \nonumber \\
& =& e^{-\frac{1}{2} ||f+jf||^2} + e^{-\frac{1}{2} ||f'+jf||^2}+ e^{-\frac{1}{2} ||f+jf'||^2} -e^{-\frac{1}{2} ||f'+jf'||^2} \;, \label{Wcccj}
\end{eqnarray}
We have  now to evaluate the scalar products between the test functions. To that purpose, we remind that, for a wedge regions, ${\cal O}={\cal W}_R$, the  modular operator $\delta$ has a continuous spectrum, given by the positive real line $[0, \infty)$ \cite{BW}. Following \cite{Summers:1987fn,Summ,DeFabritiis:2023tkh}, the test functions $(f,f')$ can be  specified as follows.  Picking up the spectral subspace of $\delta$ specified by $[\lambda^2-\varepsilon, \lambda^2+\varepsilon ] \subset (0,1)$ and introducing  the normalized vector $\phi$ belonging to this subspace, one writes 
\begin{equation}
f  = \eta  (1+s) \phi \;, \qquad f' = \eta' (1+s) i \phi \;, 
\label{nmf}
\end{equation}
where $(\eta, \eta')$ are free arbitrary parameters, corresponding to the norms of $(f,f')$.  According to the setup outlined above, equation \eqref{nmf} ensures that 
\begin{equation}
s f = f  \;, \qquad s f'= f' \;.  \label{fafa}
\end{equation}
Moreover, one checks  that $j\phi$ is orthogonal to $\phi$, {\it i.e.} $\langle \phi |  j\phi \rangle = 0$. In fact, from 
\begin{align} 
\delta^{-1} (j \phi) =  j (j \delta^{-1} j) \phi = j (\delta \phi), 
\label{orth}
\end{align}
it follows that the modular conjugation $j$ exchanges the spectral subspace $[\lambda^2-\varepsilon, \lambda^2+\varepsilon ]$ into $[1/\lambda^2-\varepsilon,1/ \lambda^2+\varepsilon ]$, ensuring that $\phi$ and $j \phi$ are orthogonal.   Concerning now the pair of Bob's test function $(jf,jf')$, we have 
\begin{equation} 
s^{\dagger}(jf) = jf \;, \qquad s^{\dagger}(jf') = jf'  \;, \label{jffp}
\end{equation}
meaning that, as required by the relativistic causality, $(jf,jf')$ belong to the symplectic  complement $\mathcal{M'}(\mathcal{O})$.  \\\\Finally, taking into account that $\phi$ belongs to the spectral subspace $[\lambda^2-\varepsilon, \lambda^2+\varepsilon ] $, it follows that \cite{Summers:1987fn,Summ,DeFabritiis:2023tkh} , in the limit $\varepsilon \rightarrow 0$:
\begin{align}
\vert\vert f \vert\vert^2  &= \vert\vert jf  \vert\vert^2 = \eta^2 (1+\lambda^2) \nonumber \\
\langle f \vert jf  \rangle &= 2 \eta^2 \lambda  \nonumber \\
\vert\vert f' \vert\vert^2  &= \vert\vert jf'  \vert\vert^2 = {\eta'}^2 (1+\lambda^2) \nonumber \\
\langle f' \vert  jf' \rangle &= 2 {\eta'}^2 \lambda  \nonumber \\
\langle f \vert jf' \rangle &=  0  \;.  \label{sfl}
\end{align}

\subsubsection{The violation of the Bell-CHSH inequality} 

We can now discuss the violation of the Bell-CHSH inequality associated to the correlation function \eqref{Wcc}. Using the scalar products of eqs.\eqref{sfl}, for $\langle 0 |\; {\cal C} \;| 0\rangle$ one gets 
\begin{equation} 
\langle 0|\;{\cal C} \;|0\rangle = e^{-\eta^2(1+\lambda)^2} + 2 e^{-\frac{1}{2}(\eta^2+ \eta'^2)(1+\lambda^2)} - e^{-\eta'^2(1+\lambda)^2}  \;. \label{vv}
\end{equation}
In order to have a concrete idea of the size of the violation achieved by eq.\eqref{vv}, one might employ the following choice 
\begin{equation} 
\eta= 0.01 \;, \qquad \eta'= 0.564058 \;, \qquad \lambda = 0.495456 \;, \label{values}
\end{equation} 
resulting in 
\begin{equation} 
\langle {\cal C} \rangle = 2.14931 \;. \label{vt}
\end{equation}
One notices that the parameters $(\eta,\eta')$, corresponding to the norms of the test functions $(f,f')$, have been treated as free parameters, akin to the Bell's angles of the Quantum Mechanical example of eq.\eqref{rc}. This possibility stems from the fact that the Weyl operators $e^{i\varphi(f)}$ and $e^{i\varphi(f')}$ are unitary for any value of $(\eta,\eta')$. Even if the result reported in expression \eqref{vt} is far from Tsirelson's bound,  $2\sqrt{2}$, it is an expressive value for the violation of the Bell inequality,  showing the role played by  all tools presented in the previous sections.

\subsection{Evaluation of the correlation functions of Hermitian  bounded  operators}\label{dich}

\begin{figure}[t!]
	\begin{minipage}[b]{0.4\linewidth}
		\includegraphics[width=\textwidth]{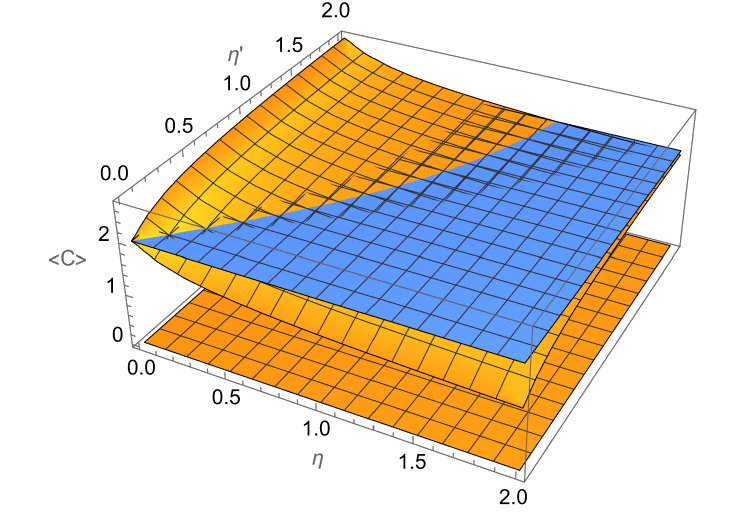}
	\end{minipage} \hfill
\caption{Plot of the Bell-CHSH correlator $\langle {\cal C} \rangle$ as a function of the parameters $(\eta, \eta')$, for the value $\lambda= 0.8$.  The violation of the Bell-CHSH inequality corresponds to the orange surface above the blue one. }
	\label{Qfig}
	\end{figure}

 Having discussed the correlation function of the Weyl operators, we move to the construction of a set of bounded Hermitian operators. 
For concreteness, let us consider the example of the bounded Hermitian operator of eq.\label{th1}
\begin{equation}  
{\tilde {\cal Q}}_h = \frac{1}{1+ \varphi(h)^2}  \;. \label{th22}
\end{equation}
 Making use of the Fourier transformation 
 \begin{equation} 
 \frac{1}{1+x^2} = \frac{1}{2}\int_{-\infty}^{\infty} dk e^{-|k|}\; e^{ikx}  \;, \label{ft}
 \end{equation}
 one writes 
 \begin{equation}  
{\tilde {\cal Q}}_h = \frac{1}{1+ \varphi(h)^2}  = \frac{1}{2} \int_{-\infty}^{\infty} dk e^{-|k|} e^{i k \varphi(h)} \; \label{qq1}
\end{equation}
Therefore, for the correlation function $\langle {\tilde {\cal Q}}_f {\tilde {\cal Q}}_{jf} \rangle$ one finds 
\begin{equation} 
\langle {\tilde {\cal Q}}_f {\tilde {\cal Q}}_{jf} \rangle = \frac{1}{4}\int dk dp e^{-|k|}e^{-|p|} \; e^{-\frac{1}{2}(k^2 \eta^2 (1+\lambda^2) + p^2\eta^2(1+\lambda^2) + 4 kp \lambda \eta^2) }
\end{equation}
 Thus, for the Bell-CHSH inequality, one gets 
 \begin{eqnarray} 
 \langle {\cal C} \rangle & = &  \langle 0|\;({\tilde {\cal Q}}_f+ {\tilde {\cal Q}}_{f'}) {\tilde {\cal Q}}_{jf}+ {\tilde {\cal Q}}_f- {\tilde {\cal Q}}_{f'}) {\tilde {\cal Q}}_{jf}  \;|0\rangle \nonumber \\
 & =& \frac{1}{4}\int dk dp e^{-|k|}e^{-|p|} \; e^{-\frac{1}{2}(k^2 \eta^2 (1+\lambda^2) + p^2\eta^2(1+\lambda^2) + 4 kp \lambda \eta^2) } \nonumber \\
 &+&  \frac{1}{2}\int dk dp e^{-|k|}e^{-|p|} \; e^{-\frac{1}{2}(k^2 \eta^2 (1+\lambda^2) + p^2\eta'^2(1+\lambda^2)  } \nonumber \\
 &-&\frac{1}{4} \int dk dp e^{-|k|}e^{-|p|} \; e^{-\frac{1}{2}(k^2 \eta'^2 (1+\lambda^2) + p^2\eta'^2(1+\lambda^2) + 4 kp \lambda \eta'^2) } \;. \label{QQ}
 \end{eqnarray}
 The violation of the Bell-CHSH inequality corresponding to eq.\eqref{QQ} is well captured by the following plot, Fig.\eqref{Qfig}
 
The orange surface gives the behavior of the Bell-CHSH correlator $\langle {\cal C}\rangle$ as a function of the parameters $(\eta, \eta')$, for $\lambda=0.8$. The blue surface is tha classical value, {\it i.e.} 2. One sees that there is a rather huge region in space parameters for which the orange surface is above the blue one. This region corresponds to the violation of the Bell-CHSH inequality. 
\section{The numerical approach}\label{na}

This last Section is devoted to the presentation of a numerical setup for the correlation function \eqref{Wcc}. As example, we shall consider the case of the real scalar massive field in $1+1$ Minkowski spacetime. \\\\Let us begin by showing the expressions of the Hadamard and of the Pauli-Jordan functions entering the scalar product between test functions 
\begin{equation} 
\langle f |g\rangle = H(f,g) + \frac{i}{2}\Delta_{PJ}(f,g) \;. \label{confsc}
\end{equation}
One has
\begin{eqnarray} 
\Delta_{PJ}(t,x) & =&  -\frac{1}{2}\;{\rm sign}(t) \; \theta \left( \lambda(t,x) \right) \;J_0 \left(m\sqrt{\lambda(t,x)}\right) \;, \nonumber \\
H(t,x) & = & -\frac{1}{2}\; \theta \left(\lambda(t,x) \right )\; Y_0 \left(m\sqrt{\lambda(t,x)}\right)+ \frac{1}{\pi}\;  \theta \left(-\lambda(t,x) \right)\; K_0\left(m\sqrt{-\lambda(t,x)}\right) \;, \label{PJH}
\end{eqnarray}
where 
\begin{equation} 
\lambda(t,x) = t^2-x^2 \;, \label{ltx}
\end{equation}
and $(J_0,Y_0,K_0)$ are Bessel functions, while $m$ is the mass parameter.  



\subsection{Expressing the Bell-CHSH inequality in terms of scalar products between test functions}

\begin{figure}[t!]
	\begin{minipage}[b]{0.4\linewidth}
		\includegraphics[width=\textwidth]{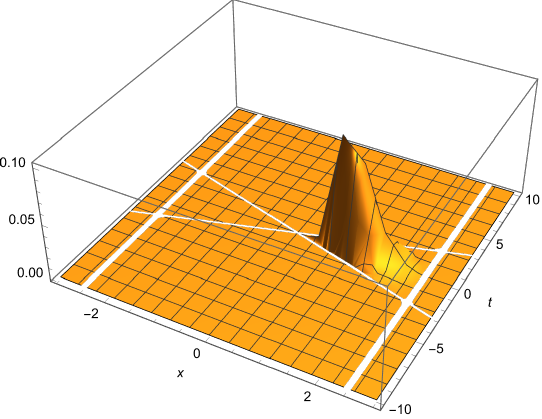}
	\end{minipage} \hfill
\caption{Plot of the test function $f(t,x)$, for $(a=1, \eta=1, \alpha=2.5 )$.   }
	\label{ff}
	\end{figure}

\begin{figure}[t!]
	\begin{minipage}[b]{0.4\linewidth}
		\includegraphics[width=\textwidth]{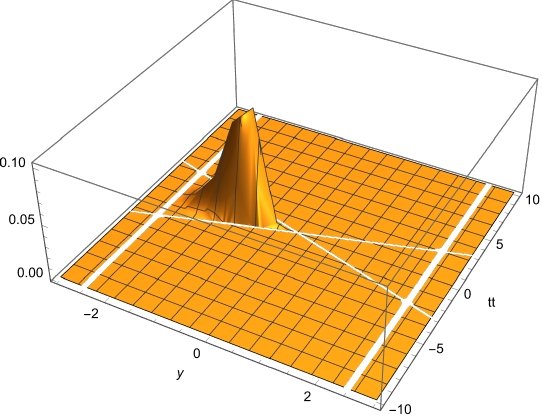}
	\end{minipage} \hfill
\caption{Plot of the test function $g(t,x)$, for $(b=1, \sigma=1, \beta=2.5 )$.   }
	\label{gg}
	\end{figure}

We proceed by specifying Alice's and Bob's test functions, respectively:  $supp(f,f') \in {\cal W}_R$ and $supp(g,g') \in {\cal W}_L$, where 
${\cal W}_R$ and ${\cal W}_L$ are the right and left wedges
\begin{equation} 
{\cal W}_R = \big\{ (x,t) \in {\mathbb{R}^2}, \; x > |t| \; \big\}\;, \qquad {\cal W}_L = \big\{ (x,t) \in {\mathbb{R}^2}, \; -x > |t| \; \big\}\;.  \label{rdw2}
\end{equation}
Explicitly, for $(f,f')$ we write 
\begin{align}
f(t,x) = \eta
\left\{
    \begin {aligned}
         & e^{-\frac{a}{x^2-t^2}}\;e^{-\frac{1}{\alpha^2-x^2}}\;e^{-(x^2+t^2)} \quad & \alpha \ge x \geq |t|  \\
         & 0 \quad & {\rm elsewhere}                   
    \end{aligned}
\right. \label{aa3}
\end{align}
and 
\begin{align}
f'(t,x) = \eta'
\left\{
    \begin {aligned}
         & e^{-\frac{a'}{x^2-t^2}}\;e^{-\frac{1}{\alpha'^2-x^2}}\;e^{-(x^2+t^2)} \quad & \alpha' \ge x \geq |t|  \\
         & 0 \quad & {\rm elsewhere}                   
    \end{aligned}
\right. \label{aa4}
\end{align}
where $(a,a',\alpha, \alpha',\eta,\eta)$ are arbitrary parameters. The  exponential factor $e^{-(x^2+y^2)}$, although not necessary for the analytic side, has been included to improve the numerical convergence of the integrals corresponding to the scalar products.  The functions $(f,f')$ are  smooth functions with support in the  right wedge ${\cal W}_R$, vanishing  at the boundary, {\it i.e.} $t=\pm x$. Analogously, in the left wedge ${\cal W}_L$, one  considers 
\begin{align}
g(t,x) = \sigma
\left\{
    \begin {aligned}
         & e^{-\frac{b}{x^2-t^2}}\;e^{-\frac{1}{\beta^2-x^2}}\;e^{-(x^2+t^2)} \quad & \beta \ge - x \geq |t|  \\
         & 0 \quad & {\rm elsewhere}                   
    \end{aligned}
\right. \label{bb3}
\end{align}
and
\begin{align}
g'(t,x) = \sigma'
\left\{
    \begin {aligned}
         & e^{-\frac{b'}{x^2-t^2}}\;e^{-\frac{1}{\beta'^2-x^2}}\;e^{-(x^2+t^2)} \quad & \beta' \ge - x \geq |t|  \\
         & 0 \quad & {\rm elsewhere}                   
    \end{aligned}
\right. \label{bb4}
\end{align}
with $(b,b',\beta,\beta',\sigma,\sigma')$ free parameters. The behavior of $f$ and $g$ is shown in Figs.\eqref{ff},\eqref{gg}. One sees that $f$ vanishes in the left wedge, while $g$ vanishes in the right one. \\\\In order to express the Bell-CHSH correlation function \eqref{Wcc} in terms of scalar products between test functions, we remind that 
\begin{align}\label{NormaFG}
			\vert\vert f+g \vert\vert^2 = H(f,f) + 2 H(f,g) + H(g,g) \;,
	\end{align}
where $H(f,g)$ stands for the multi-dimensional integral 
\begin{equation} 
H(f,g) = \int d^2x \;d^2y \;f(x) H(x-y) g(y) \;. \label{mint}
\end{equation}	
Therefore, 	
	\begin{align}\label{BellFG}
		\langle 0 \vert \mathcal{C} \vert 0 \rangle &= e^{-\frac{1}{2}\left[H(f,f) + 2 H(f,g) + H(g,g)\right]} \nonumber \\
		&+ e^{-\frac{1}{2} \left[H(f',f') + 2 H(f',g) + H(g,g)\right]} \nonumber \\
		&+ e^{-\frac{1}{2} \left[H(f,f) + 2 H(f,g') + H(g',g')\right]}  \nonumber \\
		&- e^{-\frac{1}{2} \left[H(f',f') + 2 H(f',g') + H(g',g')\right]}.
	\end{align}
Being expression \eqref{NormaFG} a norm, it follows that the Pauli-Jordan function $\Delta_{PJ}$ does not appear in equation \eqref{BellFG}. 
	
\subsection{Numerical integration and results} 
Let us give here a few details on the numerical integration of the scalar products appearing in eq.\eqref{BellFG}. The typical integral is shown in equation \eqref{mint}. Due to the difficulties of evaluating the Fourier transformation of the test functions in momentum space, expression \eqref{mint} has been evaluated as it  stands, {\it i.e.} in configuration space. We have employed Mathematica, relying on two methods of integration: QuasiMonteCarlo and MultidimensionalRule. The parameters $(a,\alpha,\eta,a', \alpha',\eta')$, $(b,\beta,\sigma, b', \beta',\sigma')$ as well as the mass $m$ have been selected by performing tests using  an aleatory algorithm. In each test,  $10^5$ aleatory values for the parameters have been checked. \\\\The following table gives an idea of the results which have been obtained. One sees that the test functions \eqref{aa3}-\eqref{bb4} give rise to a violation of the Bell-CHSH inequality. Though, as already mentioned, the concrete realization of both operators and test functions leading to the maximal violation $2\sqrt{2}$ is still an open issue.

 \begin{table}[h!]
		\begin{tabular}[t]{|c|c|c|c|c|c|c|c|c|c|c|c|c|c| }
			\hline
	$a$ & $\eta$ & $b$ & $\sigma$ & $a'$ & $\eta'$ & $b'$ & $\sigma'$ &	$\alpha$ & $\alpha'$ & $\beta$ & $\beta'$ &	$m$ &
			$\langle \mathcal{C} \rangle$  \\
			\hline 
			0.553252 & 0.501461&0.0255094 &0.0277324 &4.88226 &2.13737 & 1.13043&  6.34535 &  3.35234 &  29.6709 &  2.43472 & 39.5616 & 0.0105 & 2.036467 \\
			\hline
			 0.500578 & 0.298369 & 0.653954 & 0.0417114 & 3.61629 & 0.0116148 & 2.41375 & 13.1309 & 4.05258 & 8.10541 & 1.45682& 19.0785 & 0.0251 &2.034017 \\
			\hline
 0.61566 & 0.94915 & 0.693725 &  0.0946157 & 
    3.80309 &  1.58214 & 1.29682 & 3.46438 & 
     2.48678 & 
    148.817 &  3.18138 &  55.3358 & 
  0.00068 &   2.044862 \\
			\hline
	  0.876652 & 0.47235 &  
    0.0344563 & 0.0887357 & 
    2.92081& 0.21993 & 1.30691 &  4.7266 & 
    6.27319 &  563.98 & 
    1.46396 & 201.305 &
  0.00027 &  2.044925		 \\
			\hline
		\end{tabular}
		\caption{Results obtained for the Bell-CHSH correlation function \eqref{Wcc}. The values of the violation are reported in the last column.}
		\label{tabelaMax}
	\end{table}	
$\clubsuit$ Similar numerical results can be obtained by employing the dichotomic Hermitian operators introduced in subsection \eqref{dich}. Let us end  by mentioning that a numerical approach to the Bell-CHSH inequality in the case of the $1+1$ massless  free spinor fields can be found in \cite{Dudal:2023mij,Dudal:2024bmf} where,  unlike the scalar case,  numerical values close to the Tsirelson bound have been reported.

\section{Conclusions}\label{Cc}

In these notes we have attempted at presenting in a concise and self-contained fashion what might be called a {\it theoretical minimum} to face the Bell-CHSH inequality in Quantum Field Theory. The subject is a fascinating and challenging field, calling for a deep understanding of the relationship between entanglement and spacetime. \\\\Willing to present a short list of topics to be investigated, we could mention:
\begin{itemize} 
\item with the exception of a few lower dimensional models \cite{Summers:2003tf}, the results obtained so far for the violation of the Bell-CHSH inequality are limited to free fields. The facing of what might happen in an interacting theory is an open problem. 

\item Besides the complementary wedges $({\cal W}_R, {\cal W}_L)$, the study of a possible maximal violation of the Bell-CHSH inequality for other kinds of spacetime causal regions remains to be accomplished, see ref.\cite{DeFabritiis:2024jfy} for a recent numerical investigation. 

\item The study of the Bell-CHSH in gauge theories looks plenty of aspects to be unravelled, such as:  the role of the BRST symmetry in the constructions of the Bell-CHSH correlation function \cite{Dudal:2023pbc}, the formulation of the Bell-CHSH in non-Abelian gauge theories, the possibility of making use  of the gauge invariant Wilson loops  as bounded unitary operators, the study of the violation in presence of Higgs fields.  
\end{itemize}

\section*{Acknowledgments}
These notes have arisen from talks and lectures given by us during the last two years. We would like to express our gratitude to our friends and colleagues: J.C.A. Barata, D. Dudal, P. De Fabritiis, F. Guedes, R. Grossi e Fonseca for discussions and collaborations. I.R.  acknowledge the warm hospitality of the QIT group in ETHz where this work has been completed. The authors would like to thank the Brazilian agencies CNPq, CAPES end FAPERJ for financial support.  S.P.~Sorella, I.~Roditi, and M.S.~Guimaraes are CNPq researchers under contracts 301030/2019-7, 311876/2021-8, and 309793/2023-8, respectively. 



	
\end{document}